\documentclass[twocolumn,twocolappendix]{aastex631}

\shorttitle{LAMOST YSOs}
\shortauthors{Fang et al.}
\graphicspath{{./}{figures/}}

\begin{document}

\title{LAMOST YSOs. I. Spectroscopically identifying and characterizing M-type young stellar objects}

\correspondingauthor{Xiang-Song Fang; Jian-Rong Shi}
\email{xsfang@bao.ac.cn; sjr@nao.cas.cn}

\author[0000-0003-3240-1688]{Xiang-Song Fang}
\affil{CAS Key Laboratory of Optical Astronomy, National Astronomical Observatories, Chinese Academy of Sciences, Beijing 100101, China}

\author[0000-0002-0349-7839]{Jian-Rong Shi}
\affiliation{CAS Key Laboratory of Optical Astronomy, National Astronomical Observatories, Chinese Academy of Sciences, Beijing 100101, China}
\affiliation{School of Astronomy and Space Science, University of Chinese Academy of Sciences, Beijing 100049, China}

\author[0000-0001-6898-7620]{Ming-Yi Ding}
\affiliation{National Astronomical Observatories/Nanjing Institute of Astronomical Optics \& Technology, Chinese Academy of Sciences, Nanjing 210042, China}

\author[0000-0002-9494-0946]{Zi-Huang Cao}
\affiliation{CAS Key Laboratory of Optical Astronomy, National Astronomical Observatories, Chinese Academy of Sciences, Beijing 100101, China}



\begin{abstract}
This study utilized LAMOST low-resolution spectra to identify M-type YSOs and characterize their accretion signatures. We measured characteristic features, including hydrogen Balmer, Li {\sc i}, He~{\sc i}, Na {\sc i}, and Ca {\sc ii} lines, as well as molecular absorption bands such as CaH. These features were evaluated for their potential to distinguish between different classes of M-type stars. In addition to the commonly used H$\alpha$ emission and {Li \sc i} $\lambda6708$~\AA~absorption, other features such as He~{\sc i}, Na {\sc i}, Ca~{\sc ii} HK and IRT lines, and CN, CaH, and VO bands also show potential for identifying YSOs from M-type dwarfs and giants. Based on key red-band features, we identified over 8,500 M-type YSO candidates from the LAMOST DR8 archive using the random forest technique, with over 2,300 of them likely being classical T Tauri stars (CTTSs). By incorporating $Gaia$ astrometry, 2MASS photometry, and extinction from 3D dust maps, we estimated the basic properties (such as age and mass) of the YSO candidates based on {\sc parsec} models. We also estimated the mass accretion rates of the CTTS candidates using H$\alpha$ emission. The derived accretion rates show a dependence on age and mass similar to that described in previous studies. Significant scatter in mass accretion rates exists even among stars with similar age and mass, which may be partly attributed to variations in $\rm H\alpha$ emission and the veiling. 
\end{abstract}

\keywords{late-type stars (909) --- Stellar accretion (1578) --- Stellar spectral lines (1630) --- T Tauri stars (1681) --- Young stellar objects (1834)}


\section{Introduction} \label{sec:intro}
Young stellar objects (YSOs) are nascent stars that form through the gravitational collapse of cold interstellar material and have not yet initiated core hydrogen burning, placing them in the pre-main sequence (PMS) stage. The evolution of YSOs are complex, influenced by deep convective envelopes, strong magnetic fields, the presence of circumstellar disks, and episodic accretion and outflow processes. A comprehensive census of YSOs, particularly a well-defined, homogeneous sample of these young targets are crucial for various studies, including star and planet formation \citep{Luhman2016} and the star-forming history within the Milky Way.

YSO candidates are traditionally identified using color-color and color-magnitude diagrams, leveraging their infrared (IR) excess emissions \citep[e.g.][]{Gutermuth2008,Gutermuth2009,Rebull2010}. These emissions arise from stellar radiation reprocessed in the dusty material of their natal envelopes or circumstellar disks. However, a significant limitation of this method is the potential contamination from extragalactic sources such as PAH-rich star-forming galaxies and active galactic nuclei, which exhibit similar colors to YSOs. Recent advancements in astrometry and photometry, particularly with the $Gaia$ mission, have enabled more precise identification of YSOs. Kinematically and spatially clustering algorithms, combined with $Gaia$'s measurements of parallax and proper motions, have been used to identify YSOs in distinct star-forming regions or in the solar neighborhood \citep[e.g.,][]{Kounkel2018, Damiani2019, Galli2019, Kounkel2022, Olivares2023}. Additionally, the positions of stars on the Hertzsprung-Russell (HR) diagram have been examined to identify YSOs \citep[e.g.,][]{Zari2018}. Optical and near-infrared (near-IR) spectroscopic data also provide valuable insights, e.g., characteristic features such as evident {Li \sc i} absorption, strong H$\alpha$ and Br~$\gamma$ emissions can be used to select YSOs in nearby star-forming regions \citep[e.g.,][]{Fang2009, Connelley2010}. Elevated levels of X-ray emission, another hallmark of YSOs can also help distinguish them from field stars \citep{Feigelson1981, Feigelson1999, Winston2007}.

The current era of big data in astronomy has opened new avenues for discovering interesting targets. Machine learning techniques have been increasingly applied to large datasets to identify YSOs. For instance, \citet{Marton2019} identified over 130,000 disk-bearing YSO candidates using $Gaia$ data and $WISE$ photometry. Similarly, \citet{McBride2021} identified over 190,000 possible pre-main sequence sources and derived their ages using $Gaia$ data and 2MASS photometry with deep learning techniques. Moreover, the Large sky Area Multi-Object fiber Spectroscopic Telescope (LAMOST) has collected 10 million stellar spectra in the Milky Way, enabling the identification of stars with specific features, such as carbon stars \citep{Ji2016}, early-type emission-line stars \citep{Hou2016}, and cataclysmic variables \citep[e.g.,][]{Han2018, Hou2020}. Recent studies have also utilized LAMOST spectra to search for YSOs \citep[e.g.,][]{Zhang2024, Saad2024}, highlighting the potential of the LAMOST spectral survey in this field.

Low-mass YSOs, due to their large initial numbers and slow evolution during the PMS stage, constitute a significant portion of all YSOs. However, their faintness poses challenges for observation and further characterization. In this work, we explored the LAMOST low-resolution spectra archive to spectroscopically characterize the features of low-mass young stars and identify M-type YSOs from dwarfs and evolved stars. Given the complexity of YSO observable properties, we first employed a straightforward approach to guide our efforts and then applied machine learning techniques, specifically random forest classification, to identify YSO candidates based on red-band spectral features. We further investigated the accretion features of classical T Tauri stars (CTTSs) in our YSO sample and estimated their mass accretion rates using H$\alpha$ emission. We also conducted a statistical analysis of lithium (Li) absorptions, H$\alpha$ emissions, and investigated the scatter in mass accretion rates due to the temporal variability of H$\alpha$ emission for CTTS candidates. This comprehensive approach aims to enhance our understanding of the properties and evolution of low-mass YSOs.

\section{Data and Idenfification of YSOs}
\subsection{LAMOST Data: Pre-selection and initial sample} \label{sec:preselect}

The LAMOST, situated at Xinglong Observatory in China, is a reflecting Schmidt telescope distinguished by its wide field of view, spanning 20 $\text{deg}^{2}$ in the sky, and a substantial effective aperture of approximately 4 meters. Its focal plane hosts a total of 4000 fibers, rendering it an exceptionally efficient instrument for spectral data collection, e.g., by summer 2020, LAMOST had collected over 10.3 million stellar spectra with spectral resolving power of $R\sim1800$ (approximately 3.6~\AA~around 6500~\AA), covering a wavelength range from 3700 to 9100~\AA~\citep[see][for more details]{Zhao2006,Zhao2012,Luo2015}. For A, F, G, and K-type stars, provided the spectra meet the signal-to-noise ratio (SNR) requirement, LAMOST delivers the stellar atmospheric parameters (effective temperature, surface gravity, and metallicity) derived from the LAMOST stellar parameter pipeline, alongside heliocentric radial velocity obtained though the ULYSS \citep{Wu2011,Luo2015}. 

Given our primary interest in low-mass (M-type) YSOs, we focused on a subset of the LAMOST low-resolution spectral archive (DR8 v1.0). Our approach involved filtering out spectra of poor quality, potential non-stellar objects, and hotter stars before proceeding with the classification process. The specific steps taken were:

 (1) \textit{Signal-to-noise ratio cut}: We initially excluded spectra with an $r$-band signal-to-noise ratio (${\rm SNR}_r$) less than 8. 
 
 (2) \textit{Class cut}: We further eliminated spectra classed as `GALAXY' or `QSO' by the LAMOST 1D pipeline, which utilizes a cross-correlation method for classification \citep{Luo2015}. 
 
 (3) \textit{Spectral type cut}: M-type stars are known for their characteristic molecular band absorptions, such as TiO and CaH, which serve as effective temperature indicators \citep{Reid1995,Lepine2013,Fang2016}. We measured several spectral indices, including TiO5n and CaOH (see Appendix~\ref{sec:features} for the definitions). To select M-type spectra, we applied a cutoff of $0 < \text{TiO5n} < 0.87$, recognizing that K7- and M0-type stars typically exhibit a TiO5n value around 0.87. Moreover, to reduce contamination from hotter stars, possibly due to inaccuracies in the TiO5n region's data collection, we excluded spectra with $\text{CaOH} \geqslant 0.96$ (equivalent to a TiO5n value of approximately 0.9), given that CaOH serves as a similar temperature indicator to TiO5n.
 
 Following the application of these filters, approximately 758,600 spectra remained. We subsequently removed about 22,700 spectra of potential hotter stars (A, F, G, and early-K types) identified by the LAMOST pipeline, along with approximately 10,300 spectra with issues (bad collections) in the targeted spectral bands. Consequently, our initial sample comprised  $\sim$725,500 spectra (with a typical ${\rm SNR}_r\sim20$) representing 583,227 M-type targets (hereafter referred to as sample $\rm {S}_{0}$). Among these, 76,385 targets (13\%) possessed two-epoch spectra, while 24,152 (4\%) had at least three epochs of observations. For stars with multiple epochs of spectra, the median values (or mean values for targets with two-epoch spectra) of the relevant feature measurements were utilized.

\subsection{Identifying YSOs with LAMOST spectral features} 
The core objective of the YSO identification process is to discern spectral features that effectively differentiate among various classes. We have quantified nearly 30 potentially relevant spectral features, encompassing hydrogen Balmer lines, neutral and ionized metallic lines (such as Li~{\sc i}, Na~{\sc i}, He~{\sc i}, and Ca~{\sc ii}), certain forbidden lines ([S~{\sc ii}] and [N~{\sc ii}]), and several molecular absorption bands (TiO, CaH, and VO); many of these features exhibit a significant capacity to distinguish low-mass YSOs from M dwarf and giant stars (see Appendix~\ref{sec:features} for more details). In this study, we employed two methodologies to identify potential YSOs within the $\rm {S}_{0}$ sample: a direct, guiding method primarily utilizing Li~{\sc i} $\lambda6708$~\AA~absorption and H$\alpha$ emission lines, supplemented by two additional activity and/or gravity-related features, and a non-intuitive approach leveraging a machine learning technique, specifically the Random Forest (RF) classifier, to discern M-type YSO candidates from dwarf and giant stars. This section outlines our identifying procedures. 

\subsubsection{Lithium-filtered M-type YSOs} \label{sec:lifilter}
As elaborated in Appendix~\ref{sec:features} and illustrated in Figure~\ref{fig:cuts}, compared to M-type giants and dwarfs with comparable spectral types, YSOs exhibit stronger absorption in the Li~{\sc i} $\lambda6708$ \AA~line. In this work, we initially selected M-type YSO candidates residing within the so-called YSO region (delimited by the black solid line) in the diagram plotting the equivalent width of the Li~{\sc i} $\lambda6708$ \AA~line (EW6708) against TiO5n, as depicted in panel (a) of Figure~\ref{fig:cuts}. Note that the black line represents a smoothed fit to the EW6708 positions of the dwarf and giant samples, shifted vertically by 0.1~\AA, a margin slightly exceeding the typical uncertainty of EW6708 (with the typical $\rm {SNR}_r$ in $\rm {S}_{0}$ being around 20, a measurement error $\sim$0.08~\AA~in EW6708).  

To further minimize contamination from dwarfs and giants (e.g., Li-rich giants and young dwarfs), we incorporated three spectral features sensitive to luminosity class and/or chromospheric activity, specifically the $\lambda$8542 \AA~line (EW8542), Na~{\sc i} doublets (8172--8197~\AA, EW8190), and H$\alpha$ line (EW6563), as demonstrated in panels (b), (c), and (d) of Figure~\ref{fig:cuts}. Specifically, potential giants exhibiting stronger Ca~{\sc ii} $\lambda$8542 \AA~absorptions (below the black solid line in panel (b)) were excluded, as were potential dwarfs displaying stronger Na~{\sc i} doublet absorptions (below the black solid line in panel (c)). Additionally, targets manifesting weak H$\alpha$ emissions (below the black solid line in panel (d)) were screened out as potential dwarfs and giants. Notably, we implemented a stringent cut (the right-hand black solid line in panel (d)) to eliminate possible contamination from mid- and late-M dwarfs and giants. Ultimately, 6,300 M-type stars met these criteria and were thus identified as YSO candidates (hereafter referred to as $\rm {YSO}_{Li}$). 

\begin{figure*}
	\gridline{\fig{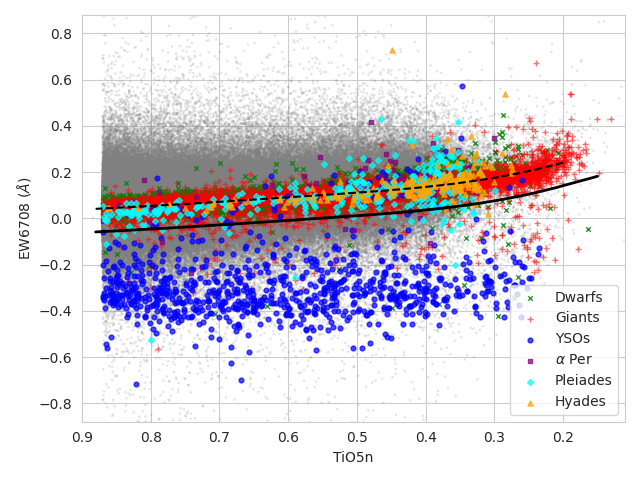}{0.5\textwidth}{(a)}
		\fig{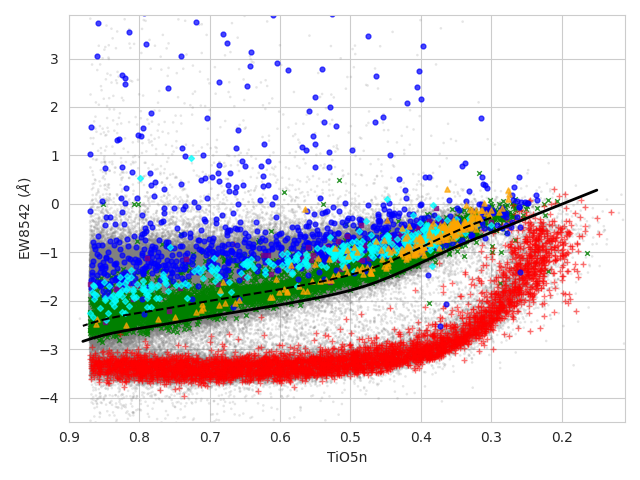}{0.5\textwidth}{(b)}}
	\gridline{\fig{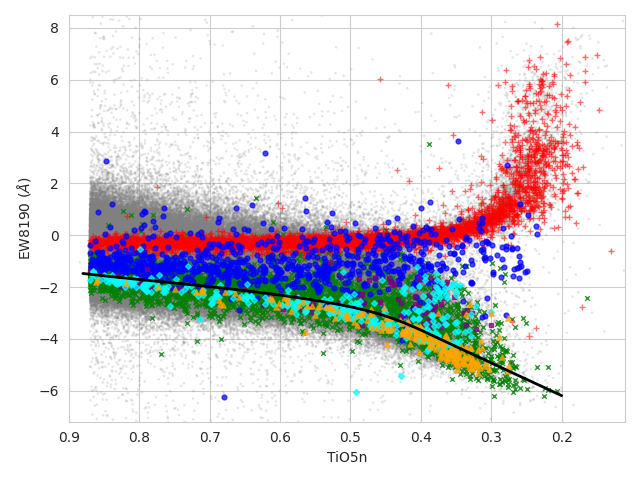}{0.5\textwidth}{(c)}
		\fig{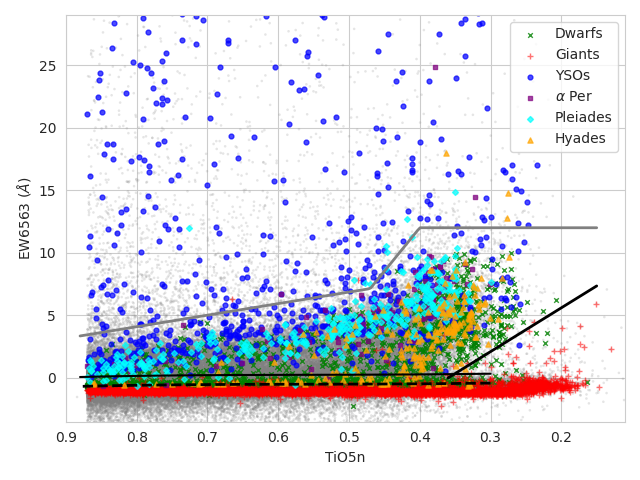}{0.5\textwidth}{(d)}}
	\caption{The equivalent widths of {Li \sc i} $\lambda6708$~\AA, {Ca \sc ii} $\lambda$8542 \AA, {Na \sc i} doublets, and H$\alpha$ line, as a function of TiO5n for the training sample stars (YSOs: blue circles, dwarfs: green crosses, giants: red pluses). The gray dots represent sources in the initial sample $\rm {S}_{0}$. For comparison, M-type members in $\alpha$ Per (purple squares), Pleiades (cyan diamonds), and Hyades (orange triangles) are also shown. (a) EW6708 against TiO5n. The black dashed line represents a smoothed fit to the mean location of M dwarfs and giants. The solid black line is a vertical shift of the dashed line by 0.1 \AA~and serves as a criterion for identifying strong lithium absorption targets. For targets with $\text{TiO5n}<0.2$, a linear extrapolation is applied. (b) EW8542 vs. TiO5n. The black dashed line is a smooth fit to the location of M dwarfs. The solid black line is a vertical shift of the dashed line by 0.32~\AA~(approximately three times the typical measurement error in EW8542) and serves as a boundary to distinguish between dwarfs and giants, aiding in the removal of potential giants. (c) EW8190 against TiO5n. The black solid line is a smooth fit to the location of M dwarfs and is used to reduce contamination from M dwarfs. (d) EW6563 vs. TiO5n. The black dashed line is a smooth fit to the location of M dwarfs, with a typical scatter of less than 0.15~\AA~around a given TiO5n value. The black solid lines (with the nearly horizontal line being a vertical shift of the dashed line by +0.75~\AA, five times the typical error in measurements of EW6563) are used to exclude chromospherically inactive dwarfs and giants. For late-M stars with $\text{TiO5n} \lesssim 0.35$, a more stringent cut (the right-hand black solid line) is applied. The gray solid line over the three open clusters serves as a cutoff to identify CTTSs (see section~\ref{sec:ctts_separation}), which exhibit H$\alpha$ emissions strong enough to exceed the saturation levels of chromospherically active young dwarfs with the same spectral types. 
		\label{fig:cuts}}
\end{figure*}

\subsubsection{Random Forest identified M-type YSOs}
The Li-filtered method for identifying YSOs may inadvertently exclude those YSOs whose surface lithium has been depleted during their early PMS phase. To address this limitation, we utilized a supervised machine learning technique, RF classifier, to identify YSO candidates based on LAMOST spectral features in the red band. In brief (see Appendix~\ref{sec:rfc} for more details), we assembled a training set encompassing three classes of M-type stars: YSOs, dwarfs, and giants. From the available measurements, we selected seven key red-band features that effectively distinguish these three classes: EW6563, EW6708, EW8190, EW8542, EW7065, CaH3n, and VO1, which include accretion tracers. Utilizing these seven features, we constructed a RF classifier composed of 400 trees to classify between M-type dwarfs, giants, and YSOs. The trained RF classifier was then applied to the initial sample $\rm {S}_{0}$, yielding a sample of $\sim$8,570 M-type YSO candidates (hereafter referred to as $\rm {YSO}_{RF}$).

\subsubsection{Separation of CTTSs}\label{sec:ctts_separation}
CTTSs are low-mass PMS stars surrounded by an accretion disk, characterized by prominent emission lines originating from infalling gas \citep{Hartmann1994}, such as the broad and strong H$\alpha$ emission line. In contrast, weak-line T Tauri stars (WTTSs) have evolved and no longer possess significant circumstellar material to accrete from (or are not currently accreting much for other reasons), typically exhibit weaker and narrower H$\alpha$ emission lines, which are not indicative of active accretion but rather attributed to their magnetically active chromospheres. It is well-established that chromospheric H$\alpha$ emission is limited by the saturation level of the active chromosphere, as observed in rapidly rotating, low-mass stars \citep[e.g.,][]{Newton2017,Fang2018}. Consequently, H$\alpha$, being one of the strongest emission lines in CTTSs, has been extensively studied as an indicator of accretion \citep[e.g.,][]{Muzerolle2003,Natta2004,Oliveira2009} and is commonly used to distinguish CTTSs from WTTSs \citep[e.g.,][]{BarradoyNavascues2003,White2003,SzegediElek2013}.

In this study, we distinguished CTTSs from WTTSs solely based on the strength of H$\alpha$ emission, rather than employing other accretion tracers. Specifically, an object was classified as a CTTS if its EW6563 value surpassed the saturation limit of chromospheric activity. Our saturation level of chromospheric activity was calibrated using M-type members of the $\alpha$ Per, Pleiades, and Hyades clusters, as most M-type stars in these young open clusters are believed to be in the chromospheric saturation region \citep[e.g.,][]{Fang2018}. Specifically, as shown by the solid gray line in panel (d) of Figure~\ref{fig:cuts}, the EW6563 values for stars with TiO5n $\gtrsim 0.47$ were determined by vertically shifting the linear fit of $\alpha$ Per and Pleiades members by 3~\AA. This 3~\AA~shift is approximately 2.5 times the scatter of EW6563 around members in these two open clusters with similar TiO5n values. For late-M members in these three clusters with $\rm TiO5n < 0.4$, most (98 percent) have EW6563 values less than 12~\AA, so we set this value as the cutoff. In the transitional interval of $0.4 \leqslant \rm TiO5n < 0.47$, a linear line connecting these two regions was determined. Overall, fewer than 3 percent of M-type members in the $\alpha$ Per and Pleiades samples lie above the gray solid line, likely due to flare events, indicating a reasonable baseline for H$\alpha$ emission representing the combination of ongoing accretion and saturated chromospheric activity. We identified over 2,300 M-type CTTS candidates whose EW6563 values exceed the specified chromospheric saturation level, suggesting they are likely in the accreting phase.

\subsection{Subdivision of $\rm {YSO}_{RF}$ sample} \label{sec:valid2}
The $\rm {YSO}_{Li}$ candidates were identified using a simple, arbitrary lithium filter along with three additional loose filters, as illustrated in Figure~\ref{fig:cuts}. This approach may exclude YSOs that have undergone lithium depletion at earlier ages and could include non-YSOs such as young dwarfs. 
Indeed, $\rm {YSO}_{Li}$ sample contains 6,300 targets showing detected {Li \sc i} $\lambda6708$ \AA~absorptions, with about 80\% of them being identified as $\rm {YSO}_{RF}$ candidates (over 19\% and less than 1\% identified as dwarfs and giants, respectively, by the RF method). Conversely, the $\rm {YSO}_{RF}$ sample includes many targets with weak or undetected {Li \sc i} $\lambda6708$ \AA~absorptions.

\begin{figure*}
	\gridline{\fig{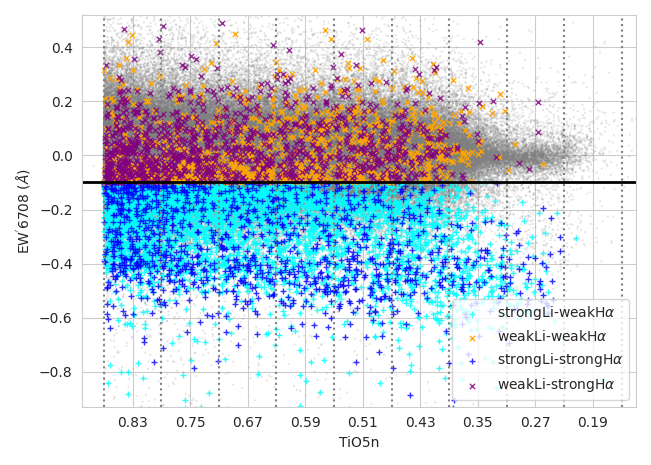}{0.5\textwidth}{}
		\fig{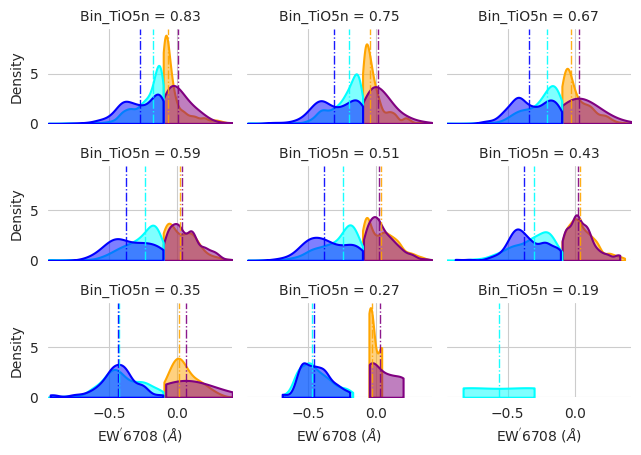}{0.5\textwidth}{}		    
	}
	\gridline{\fig{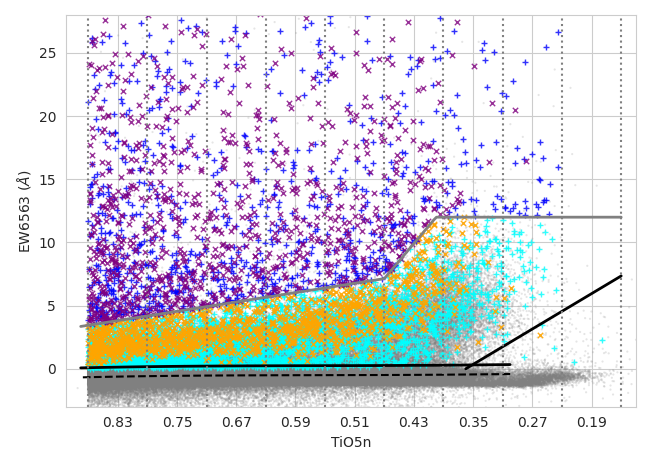}{0.5\textwidth}{}
		\fig{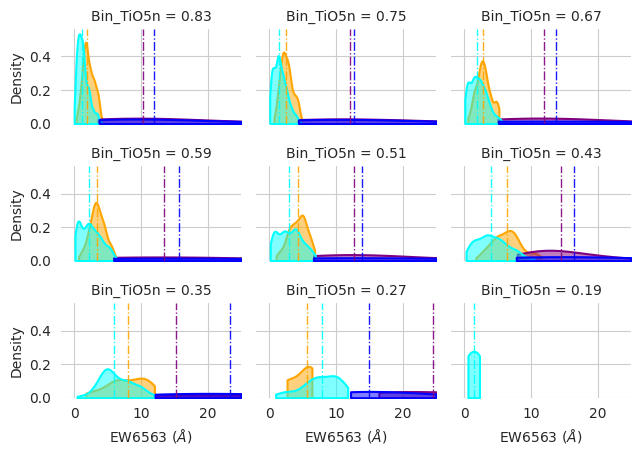}{0.5\textwidth}{}		    
	}
	\caption{Classification of $\rm {YSO}_{RF}$ candidates into four subgroups based on Li absorption and H$\alpha$ emission. Upper left: The excess equivalent width ($\rm EW^{'}6708=EW6708-EW6708_{basal}$) of {Li \sc i} $\lambda6708$, corrected for influences from nearby continua and potential contamination from adjacent lines using the reference baseline EW6708$_{\rm basal}$ (indicated by the black dashed line in panel (a) of Figure~\ref{fig:cuts}), versus TiO5n. The Li-based subgroup classification is defined at $\rm EW^{'}6708 = -0.1$~\AA~(black solid line), distinguishing strong-Li subgroups (pronounced Li absorption) from weak-Li subgroups, which exhibit undetected or negligible Li absorption centered around $\rm EW^{'}6708 = 0$~\AA. However, substantial scatter exists in $\rm EW^{'}6708$ for weak-Li subgroups, with some targets showing apparent lithium emission features. A comparison of spectral properties reveals that targets with $\rm EW^{'}6708 > 0.2$~\AA~have spectra with median $\rm {SNR}_r\sim10$ and EW6708 measurement uncertainties around 0.17~\AA, whereas those with $-0.1 < \rm EW^{'}6708 < 0.1$ \AA~have higher $\rm {SNR}_r$ (median $\sim16$) and lower EW6708 uncertainties (median $\sim0.10$~\AA). A further visual inspection of spectra for $\rm EW^{'}6708 > 0.2$~\AA~targets indicates that the lithium emission-like features are predominantly attributable to high noise levels and/or data artifacts rather than astrophysical origins. Lower left: Reproduction of panel (d) from Figure~\ref{fig:cuts}, tailored to $\rm {YSO}_{RF}$ candidates, with H$\alpha$-subgroups delineated by the gray solid line. Symbol conventions: Purple crosses (weak-Li-strong-H$\alpha$), blue pluses (strong-Li-strong-H$\alpha$), orange crosses (weak-Li-weak-H$\alpha$), cyan pluses (strong-Li-weak-H$\alpha$). Small gray dots represent the initial samples. The dotted gray lines indicate the borders of TiO5n bins. Right panels: Distributions of $\rm EW^{'}6708$ and EW6563 within each TiO5n bin, with dotted-dashed lines representing the individual medians.
		\label{fig:YrfHaLia}}
\end{figure*}

\begin{figure}
	\gridline{\fig{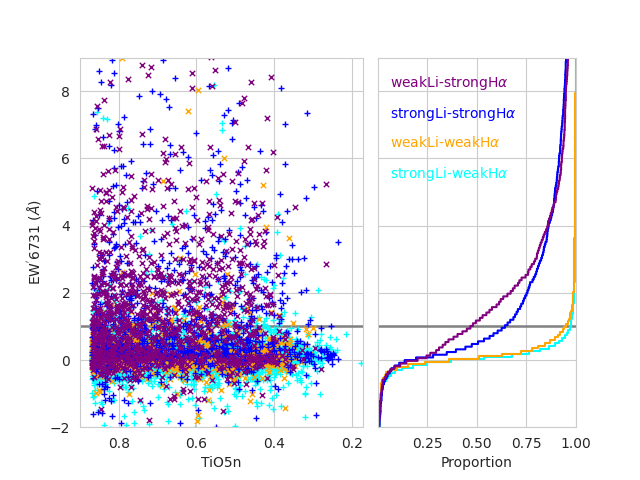}{0.5\textwidth}{}}
	\caption{The distribution of excess equivalent width for the [S~{\sc ii}] $\lambda6731$ \AA~line among $\rm YSO_{RF}$ candidates. The excess equivalent width is defined as $\rm EW^{'}6731 = EW6731-EW6731_{basal}$, where $\rm EW6731_{basal}$ represents the basal reference value determined as the mean EW6731 of M-type dwarfs and giants in the training sample with comparable spectral subtypes to the target object. This method accounts for potential influences from the nearby continuum and from absorptions due to adjacent lines, such as Fe~{\sc i}. The left panel employs symbols and color coding consistent with those in Figure~\ref{fig:YrfHaLia}, while the right panel provides the corresponding cumulative distribution functions to emphasize the differences between the various subgroups. The solid gray line at $\rm EW^{'}6731=1$~\AA~, approximately five times the typical measurement error in EW6731, serves as an empirical threshold for detecting emission in this forbidden line. Notably, approximately 45\% of strong H$\alpha$ $\rm {YSO}_{RF}$ candidates (54\% of purple crosses, and 37\% of blue pluses) have $\rm EW^{'}6731>1$~\AA, while only a small fraction ($\lesssim4$\%) of weak H$\alpha$ $\rm {YSO}_{RF}$ candidates exhibit $\rm EW^{'}6731>1$~\AA. See section~\ref{sec:valid2} for further details.
    \label{fig:ew6731}}
\end{figure}

To gain a deeper understanding of the $\rm {YSO}_{RF}$ sample in terms of Li absorption and H$\alpha$ emission, we divided them into four groups, as illustrated by Figure~\ref{fig:YrfHaLia}: 

(1) Weak Li absorption and strong H$\alpha$ emission (weak-Li-strong-H$\alpha$): This group contains 1092 (12.7\%) $\rm {YSO}_{RF}$ candidates. As shown in the right panels of Figure~\ref{fig:YrfHaLia}, these stars exhibit the weakest {Li \sc i} $\lambda6708$ \AA~absorptions and very strong H$\alpha$ emissions. Their strong H$\alpha$ emissions, exceeding the saturation level of chromospheric activity, suggest an additional contribution from processes such as accretion activity, classifying them as CTTS candidates. Examination of forbidden emission lines such as [N~{\sc ii}] $\lambda6583$ \AA~and [S~{\sc ii}] $\lambda6731$ \AA~revealed that over half (54\%) of them show evident emissions with $\rm EW^{'}6731>1$~\AA, as shown in Figure~\ref{fig:ew6731}, indicating the presence of star-disk interaction environments (e.g., accretion-powered outflows or winds).

(2) Strong Li absorption and strong H$\alpha$ emission (strong-Li-strong-H$\alpha$): This group comprises 1,236 (14.4\%) $\rm {YSO}_{RF}$ candidates. They exhibit the strongest {Li \sc i} $\lambda6708$ absorptions and the strongest H$\alpha$ emissions, as shown in right panels of Figure~\ref{fig:YrfHaLia}. The strong H$\alpha$ emission are likely due to accretion activity, classifying these stars as CTTS candidates, as discussed in Section~\ref{sec:ctts_separation}. Over a third (37\%) of these stars show evident emissions in forbidden lines such as [N~{\sc ii}] $\lambda6583$ \AA~and [S~{\sc ii}] $\lambda6731$ \AA, with $\rm EW^{'}6731>1$~\AA, as shown in Figure~\ref{fig:ew6731}. 

(3) Weak Li absorption and weak H$\alpha$ emission (weak-Li-weak-H$\alpha$): This group includes 1,242 (14.5\%) $\rm {YSO}_{RF}$ candidates. Overall, they have stronger {Li \sc i} $\lambda6708$ \AA~absorptions than the weak-Li-strong-H$\alpha$ group and slightly stronger H$\alpha$ emissions than the strong-Li-weak-H$\alpha$ group, as shown in Figure~\ref{fig:YrfHaLia}. Given their weak H$\alpha$ emissions, they are likely WTTSs. Indeed, most members of this group do not show evident emissions in the [S~{\sc ii}] $\lambda6731$ \AA~line, with only 6\% having $\rm EW^{'}6731>1$~\AA, as shown in Figure~\ref{fig:ew6731}. 

(4) Strong Li sbsorption and weak H$\alpha$ emission (strong-Li-weak-H$\alpha$): This group contains 4,997 (58.3\%) $\rm {YSO}_{RF}$ candidates. Based on their H$\alpha$ emissions, they are likely WTTS candidates. Compared to the strong-Li-strong-H$\alpha$ group, they have slightly weaker {Li \sc i} $\lambda6708$ \AA~ absorptions and even weaker H$\alpha$ emissions than the weak-Li-weak-H$\alpha$ group, as shown in Figure~\ref{fig:YrfHaLia}. Only a very small fraction (3\%) of this group show evident emissions in the  [S~{\sc ii}] $\lambda6731$ \AA~line, with $\rm EW^{'}6731>1$~\AA, as shown in Figure~\ref{fig:ew6731}. 


\subsection{Validation in YSO Identification}
\begin{figure*}
	\gridline{\fig{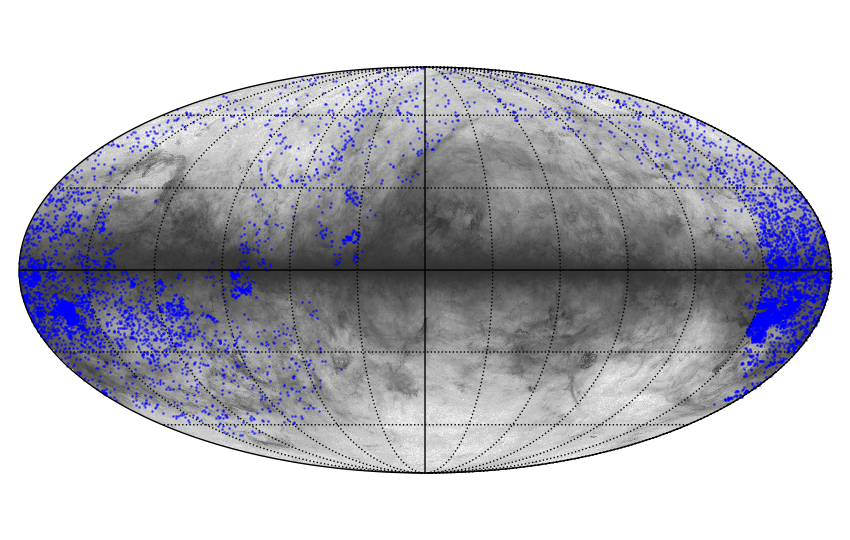}{0.5\textwidth}{a}
		\fig{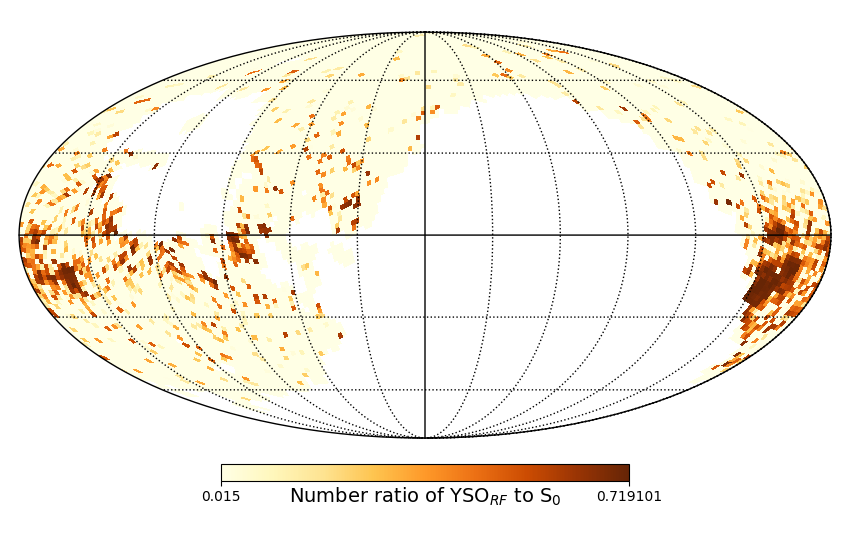}{0.5\textwidth}{b} 
	}
	\caption{(a) Sky distribution of identified YSO candidates (Mollweide projection in Galactic coordinates with $l = b = 0$ at the center, north up, and $l$ increasing from right to left). Blue dots $\rm {YSO}_{RF}$ represent candidates, and the background shows the Planck dust emission map (857 GHz) \citep{PlanckCollaboration2020}. (b) The number ratio of $\rm {YSO}_{RF}$ candidates to $\rm {S}_{0}$ sources in each sky pixel (the resolution of the sky-pixelization is approximately 110 arcmin). Lighter colors indicate lower number ratios of YSO candidates, highlighting the sky distribution of $\rm {S}_{0}$ sources. Note that the YSO candidates primarily concentrate in two regions: the Taurus \& Perseus complex (left-hand side around $l\sim173$ deg, $b\sim-15$ deg), and the Orion Molecular Cloud complex (right-hand side around $l\sim209$ deg, $b\sim-19$ deg).  \label{fig:projection}}
\end{figure*}

Figure \ref{fig:projection} shows the Galactic distribution of the identified YSO candidates. Notably, the YSO candidates predominantly cluster in two major sky regions: the Orion Molecular Cloud complex ($l\sim209^\circ$, $b\sim-19^\circ$), and the Taurus and Perseus star-forming region (including the Perseus star-forming region and the Taurus-Auriga complex, $l\sim173^\circ$, $b\sim-15^\circ$). Other recognizable grouped structures include Serpens ($l\sim30^\circ$, $b\sim+04^\circ$), Cygnus ($l\sim85^\circ$, $b\sim0^\circ$), and NGC 2264 ($l\sim203^\circ$, $b\sim+02^\circ$).

\subsubsection{Validation with star-forming regions}
To validate and compare the performance of our methods for identifying M-type YSOs, we utilized two well-known star-forming regions: the Orion Complex and the Perseus star-forming region. 

\begin{figure}
	\gridline{\fig{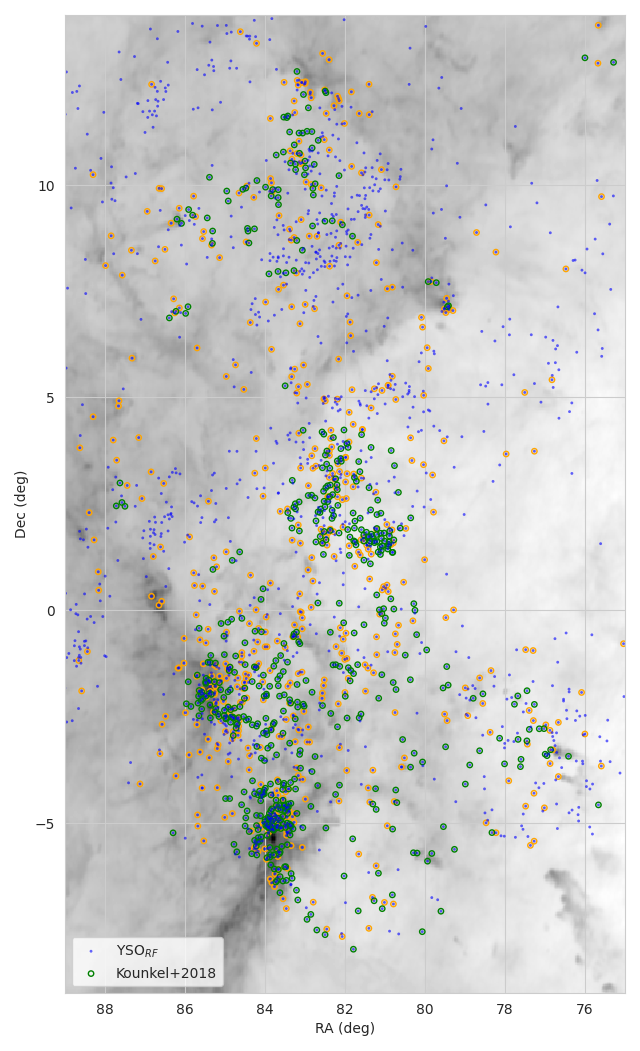}{0.5\textwidth}{}}
	\caption{Sky distribution of $\rm YSO_{RF}$ candidates (blue dot) toward the Orion star-forming region. Known dynamically grouped stars from \citet{Kounkel2018} are superimposed with green open circles. The orange open circles represent 531 possible members that are not included in the sample of \citet{Kounkel2018} but share similar distance and proper motions with Orion complex members, specifically with $-4.0<\mu_{\alpha}<4.0~\rm mas~yr^{-1}$ , $-4.0<\mu_{\delta}<4.0~\rm mas~yr^{-1}$, and distances ranging from 285 to 500 pc. The background image shows the Planck dust emission map (857 GHz).
  \label{fig:ra_dec_orion}}
\end{figure}

\begin{figure}
	\gridline{\fig{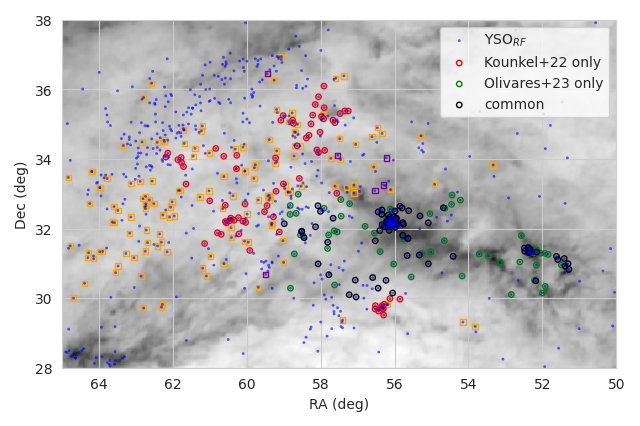}{0.5\textwidth}{}}
	\caption{Sky distribution of $\rm {YSO}_{RF}$ candidates (blue dot) toward the Perseus star-forming region. Known dynamically grouped stars are superimposed with open circles: green circles are uniquely included in the sample of \citet{Olivares2023}, red circles are only in the sample of \citet{Kounkel2022}, and black circles are common to both samples. Open squares represent possible members not included in the samples of \citet{Olivares2023} and \citet{Kounkel2022} but share similar distance and proper motions with the Per OB2a and Per OB2b groups \citep{Kounkel2022}. Orange squares (142 stars) correspond to Per OB2a, with proper motions $2.0<\mu_{\alpha}<5.6~\rm mas~yr^{-1}$ and $-7.8<\mu_{\delta}<-3.5~\rm mas~yr^{-1}$, and distances ranging from 300 to 420 pc. Purple squares (6 stars) correspond to Per OB2b, with proper motions  $5.6<\mu_{\alpha}<9.0~\rm mas~yr^{-1}$ and $-11.0<\mu_{\delta}<-7.8~\rm mas~yr^{-1}$, and distances ranging from 270 to 305 pc. The background image shows the Planck dust emission map at 857 GHz. \label{fig:ra_dec_perseus}}
\end{figure}

\citet{Kounkel2018} identified spatially and kinematically distinct groups of YSOs with ages ranging from 1 to 12 Myr in the Orion Complex by applying a hierarchical clustering algorithm to five-/six-dimensional data from APOGEE-2 and Gaia DR2. Based on the membership assignments in their table 2, we retrieved 593 stars from our initial sample $\rm {S}_{0}$. Of these 593 stars, 519 and 561 were successfully identified as $\rm YSO_{Li}$ and $\rm YSO_{RF}$ candidates, respectively, indicating recovery rates of approximately 88\% and 95\%. These results are summarized in Table~\ref{tab:vad_ori} and visualized in Figure~\ref{fig:ra_dec_orion}. Recently, \citet{Olivares2023} identified over 1,000 candidate members with ages between 3 and 10 Myr in seven Perseus physical groups by applying Bayesian methodologies to the public $Gaia$, APOGEE, 2MASS, and Pan-STARRS catalogs. A total of 158 candidate members from \citet{Olivares2023} were found in our sample $\rm {S}_{0}$, of which 132 and 149 were successfully recovered as $\rm YSO_{Li}$ and $\rm YSO_{RF}$ candidates, respectively, indicating recovery rates of approximately 84\% and 94\%. These results are detailed in Table~\ref{tab:vad_per} and illustrated in Figure~\ref{fig:ra_dec_perseus}. Additionally, based on $Gaia$ EDR3 data, \citet{Kounkel2022} identified several young groups in the same but much broader sky region toward Per OB2 (see Figure~\ref{fig:ra_dec_perseus}). Out of the 169 candidate members of these young groups present in our sample $\rm {S}_{0}$, 143 (85\%) and 159 (94\%) were successfully recovered as $\rm YSO_{Li}$ and $\rm YSO_{RF}$ candidates, respectively, demonstrating similar recovery rates. 


\subsubsection{Contamination analysis}
To evaluate the extent of contamination by particularly young dwarf stars in our YSO identification procedures, we applied both the Li- and RF-methods to members of three nearby benchmark open clusters: 
$\alpha$ Per, Pleiades, and Hyades (see Appendix~\ref{sec:ocs} for details). Among the 95 members of $\alpha$ Per in our initial sample $\rm {S}_{0}$, 3 (3.1\%) and 5 (5.2\%) were classified as $\rm {YSO}_{Li}$ and $\rm {YSO}_{RF}$ candidates, respectively. For the 362 Pleiades members in $\rm {S}_{0}$, 15 (4.1\%) and 10 (2.8\%) were identified as $\rm {YSO}_{Li}$ and $\rm {YSO}_{RF}$ candidates, respectively. As expected, none of the 218 Hyades members in $\rm {S}_{0}$ were identified as YSOs using either method. Therefore, the contamination from young M dwarfs is not significant and is thus tolerable. However, it is important to note that M-type stars, especially mid- and late-M type, with ages around 100 Myr are still in the PMS phase or just reaching the zero-age main sequence. These low YSO detection rates in Per and Pleiades suggest that both methods are less effective at identifying older PMS stars.

Another potential source of contamination in YSO prediction comes from M-type evolved stars. We examined the location of our YSO candidates with valid $Gaia$ parameters ($\varpi/\sigma_{\varpi}>5$, $\rm RUWE<1.4$) in the absolute J-band magnitudes ($M_{\rm J}$) versus $T_{\rm eff}$ (see Figure~\ref{fig:cmd_parsec}). We found that 44 ($\sim0.7\%$) of the 6,660 $\rm {YSO}_{RF}$ candidates and eight ($\sim0.2\%$) of the 4893 $\rm {YSO}_{Li}$ candidates are located in the region with $M_{\rm J}>0$~mag, which is characteristic of M-type giants. Among these 44 $\rm {YSO}_{RF}$ candidates in the M giant region, 24 targets have SIMBAD type tags, with 20 being evolved stars (11 carbon stars, 7 long-period variables, and 2 Mira stars). Further cross-matching all $\rm {YSO}_{RF}$ targets with SIMBAD revealed that 97 targets ($\sim1.1\%$) have tags indicating they are carbon stars or giants (35 carbon stars, 40 long-period variables, 19 Mira stars, and 3 AGB stars). Furthermore, cataclysmic variables (CVs) often exhibit prominent emissions in hydrogen Balmer and He~{\sc i} lines \citep[e.g.,][]{Hou2020}, sharing some spectral features with YSOs. Cross-matching our $\rm YSO_{RF}$ sample with SIMBAD database identified 27 ($\sim0.3\%$) targets labeled as CVs. 

\begin{deluxetable*}{lrrrrrr}
	\tablecaption{Comparison of our YSO candidates to members in Orion groups idtenfied by \citet{Kounkel2018}  \label{tab:vad_ori}}
	\tablewidth{0pt}
	\tablehead{
		\colhead{Group Name} & \colhead{In $\rm {S}_{0}$} & \colhead{Li-recovered}   & \colhead{RF-recovered} & \colhead{Li-recovery rate (\%)} & \colhead{RF-recovery rate (\%)}
	}
	\startdata
	ONC              & 103   & 96    & 102    & 93   & 99  \\
	L 1641N          & 8     & 8     & 8      & 100  & 100 \\
	NGC 2024         & 21    & 19    & 19     & 90   & 90  \\
	Ori B-N          & 4     & 4     & 4      & 100  & 100 \\
	$\sigma$ Ori     & 31    & 31    & 31     & 100  & 100 \\
	Ori C-C          & 40    & 33    & 36     & 83   & 90  \\
	Ori C-N          & 11    & 8     & 11     & 73   & 100 \\
	$\psi^{2}$ Ori   & 56    & 51    & 54     & 91   & 96  \\
	25 Ori           & 63    & 50    & 56     & 79   & 89  \\
	22 Ori           & 16    & 12    & 16     & 75   & 100 \\
	$\delta$ Ori     & 7     & 6     & 6      & 86   & 86  \\ 
	$\epsilon$ Ori   & 20    & 17    & 19     & 85   & 95  \\
	$\zeta$ Ori      & 28    & 26    & 27     & 93   & 96  \\
	$\eta$ Ori       & 22    & 21    & 20     & 95   & 91  \\
	Ori D-S          & 9     & 8     & 9      & 89   & 100 \\
	L 1616           & 23    & 16    & 22     & 70   & 96  \\	
	NGC 1981         & 26    & 20    & 22     & 77   & 85  \\	
	Ori X            & 20    & 17    & 20     & 85   & 100 \\
	Rigel            & 4     & 3     & 4      & 75   & 100 \\
	$\lambda$ Ori    & 30    & 27    & 27     & 90   & 90  \\  
	B30              & 21    & 19    & 19     & 90   & 90  \\
	B35              & 19    & 18    & 18     & 95   & 95  \\
	spurious         & 11    & 9     & 11     & 82   & 100 \\
	\hline                                           
	total            & 593 (553$^a$, 40$^b$)   & 519 (519, 0)   & 561 (531, 30)    & 88 (94, 0)  & 95 (96, 75)  \\
	\enddata
	\tablecomments{$a$: targets with detected {Li \sc i} $\lambda6708$ \AA~absortions, classified as strong-Li group stars in section~\ref{sec:valid2}. $b$: targets without detected {Li \sc i} $\lambda6708$ \AA~absortions, classified as weak-Li group stars in section~\ref{sec:valid2}. Other numbers in parenthesis in the last row have the same meanings. }
\end{deluxetable*}

\begin{deluxetable*}{lrrrrrr}
	\tablecaption{Comparison of our YSO candidates to members in Perseus groups identified by \citet{Olivares2023}  \label{tab:vad_per}}
	\tablewidth{0pt}
	\tablehead{
		\colhead{Group Name} & \colhead{In $\rm {S}_{0}$} & \colhead{Li-recovered}   & \colhead{RF-recovered} & \colhead{Li-recovery rate (\%)} & \colhead{RF-recovery rate (\%)}
	}
	\startdata
	NGC 1333         & 15  & 13     & 15    & 87     & 100 \\
	Autochthe        & 7   & 7      & 7     & 100    & 100 \\
	IC 348 core      & 38  & 37     & 38    & 97     & 100 \\
	IC 348 halo      & 25  & 24     & 25    & 96     & 100 \\
	Heleus           & 23  & 18     & 21    & 78     & 91  \\
	Gorgophone core  & 8   & 7      & 7     & 88     & 88  \\
	Gorgophone halo  & 29  & 23     & 27    & 79     & 97  \\
	Alcaeus          & 13  & 3      & 9     & 23     & 69  \\
	\hline
	total            & 158 (139, 19) & 132 (132, 0)    & 149 (136, 13)   & 84 (95, 0)    & 94 (98, 68)\\
	\enddata
\tablecomments{The numbers in parenthesis in the last row have the same meanings as those in Table~\ref{tab:vad_ori}. }
\end{deluxetable*}

\section{Stellar properties and accretion signatures}
\subsection{Stellar properties of YSO candidates}
\begin{figure*}
	\gridline{\fig{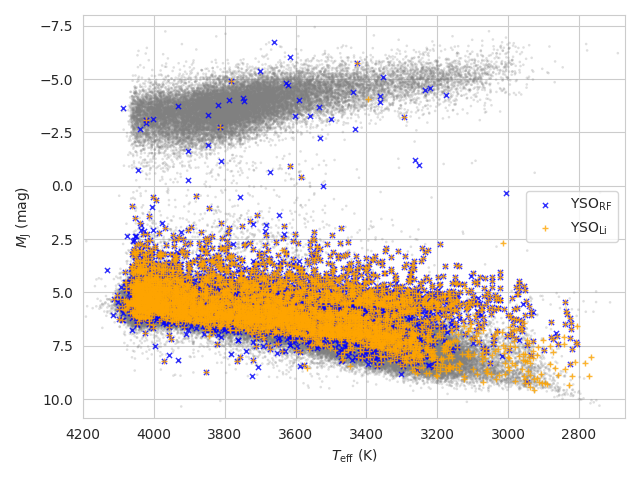}{0.5\textwidth}{}
		\fig{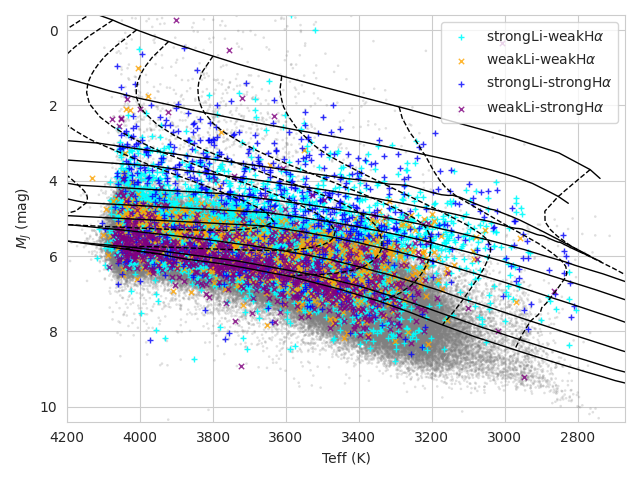}{0.5\textwidth}{}}
	\caption{Left: diagram of $M_{\rm J}$ versus $T_{\rm eff}$ for 4893 $\rm YSO_{Li}$ candidates and 6660 $\rm YSO_{RF}$ candidates with reliable $Gaia$ parallax measures ($\varpi/\sigma_{\varpi}>5$, RUWE$<$1.4). Right: zoomed-in view of the left panel, focusing on $\rm YSO_{RF}$ divided into four groups (see section~\ref{sec:valid2}). Evolutionary models from {\sc parsec} with solar metallicities are superimposed, where black dashed lines represent evolutionary tracks (from right to left: 0.1, 0.2, 0.3, 0.4, 0.5, 0.6, 0.7, and 0.75 M$_{\odot}$), and black solid lines represent isochrones (from top to bottom: 0.01, 0.1, 1, 2, 5, 10, 20, 50, 100, and 200 Myr). 
		\label{fig:cmd_parsec}}
\end{figure*}

To determine the stellar properties (such as the masses, radii, and ages) of these young stars, we utilized the evolutionary models from {\sc parsec} \citep[v1.2S,][]{Chen2014} with solar metallicity. We applied these models to a diagram of $M_{\rm J}$ (2MASS $J$-band absolute magnitude) against $T_{\rm eff}$ (the effective temperature), as shown by right panel of Figure~\ref{fig:cmd_parsec}. In fact, several sets of stellar evolutionary models can be used to estimate stellar properties, and systematic offsets may exist between them \citep[e.g.,][]{Hillenbrand2007, Herczeg2015}. In this study, we chose the {\sc parsec} isochrones due to their comprehensive coverage of the parameter space relevant to our low-mass sample stars. 

For the majority of our sample stars, effective temperatures ($T_{\rm eff}$) were derived from \citet{Ding2022},  who estimated them by comparing LAMOST low-resolution red-band spectra against model spectra produced with the MILES interpolator. Where these estimates were not available, we translated spectral types into $T_{\rm eff}$ using the conversion method detailed by \citet{Fang2016}. For stars with multiple spectra,  we adopted the median $T_{\rm eff}$ values (or the mean value for targets with exactly two epochs of data).
	
It should be noted that, as highlighted by \citet{Ding2022}, there might be systematic biases in the determination of $T_{\rm eff}$, e.g., for M dwarfs, where an underestimation of $\Delta T_{\rm eff}=-176$ K relative to ASPCAP values has been reported, whereas the discrepancy is smaller at $\Delta T_{\rm eff}=-88$ K when compared to other studies. \citet{Ding2022} do not provide a comparable assessment for M-type PMS stars. Estimating $T_{\rm eff}$ for PMS stars is notably challenging due to their often spotted surfaces caused by strong magnetic activity, exemplified by the extensive dark-spot coverages observed on WTTS stars~\citep{GullySantiago2017,PerezPaolino2024}. CTTSs add another layer of complexity, characterized by additional features such as excess continuum and emission lines from active accretion processes, and potential thermal contributions from inner disk regions \citep[e.g.][]{PerezPaolino2025}. Considering these factors, and for simplicity in this study, we adopt an uncertainty of 200 K for the $T_{\rm eff}$ of all our YSO candidates.

The choice of the $J$-band magnitude is a compromise between the availability of 2MASS data, the impact of hot accretion shocks (UV excess), and the influence of circumstellar disks (infrared excess), as well as the effects of extinction. The $J$-band is less affected by extinction and UV excess compared to shorter wavelengths, and the emission from circumstellar dust is less prominent than at longer infrared wavelengths \citep{Kenyon1990}, although the the $J$-band is still influenced by infrared excess \citep[e.g.,][]{Cieza2005}.

The absolute magnitude $M_{\rm J}$ was derived from the 2MASS $J$-band apparent magnitude $J$, the distance $d$, and the extinction in $J$-band (reddening $A_{\rm J}$). The extinction $A_{\rm J}$ was estimated using the 3D dust map of Bayestar19 \citep{Green2019}, and the distance $d$ was derived from $Gaia$ DR3 parallax \citep{GaiaCollaboration2016,GaiaCollaboration2023}. The adopted $T_{\rm eff}$, $J$, $A_{\rm J}$, $d$, and derived masses, raddi, and ages from {\sc parsec} models are listed in Table~\ref{tab:properties}.

To evaluate the intrinsic uncertainties in the determination of age, mass, and radius for our sample stars using the method described above, we employed a Monte Carlo approach. Specifically, we conducted 500 Monte Carlo simulations for each target, assuming that $T_{\rm eff}$, $J$, $A_{\rm J}$, and $d$ follow normal distributions with standard deviations $\sigma(T_{\rm eff})$, $\sigma(J)$, $\sigma(A_{\rm J})$ and $\sigma(d)$, respectively. The errors of $J$ provided by 2MASS were adopted as $\sigma(J)$, with typical values of $\sigma(J) \lesssim 0.03$ mag. The distance uncertainty $\sigma(d)$ was estimated from the parallax errors given by $Gaia$ DR3, typically resulting in $\sigma(d)/d\sim 0.02$, which corresponds to an uncertainty contribution of about 0.04 mag in the determination of $M_J$. For simplicity, $\sigma(T_{\rm eff})$ was set to be 200 K, and the standard deviation of the reddening in the $J$-band, $\sigma(A_J)$, was set to 0.08 mag for all sample stars, considering that the reddening uncertainty is typically less than 0.1 mag in $E(g-r)$ \citep{Green2019}. The simulations show that the internal uncertainty in age determination is approximately 2 Myr for stars with ages of $\sim$10 Myr, but a larger value for older stars, e.g., typically a relative uncertainty $\Delta t/t \lesssim 0.5$, mainly resulted from the large uncertainty in effective temperature. The typical uncertainties in mass and radius are 0.06 $M_{\odot}$ and 0.08 $R_{\odot}$, respectively. The simulations also reveal that the primary source of mass uncertainty is the error in $T_{\rm eff}$, rather than the uncertainty in $M_J$ determination (which is influenced by the uncertainties in $J$, $A_J$, and $d$). This is expected, as the mass evolutionary tracks for our sample stars are approximately perpendicular to the $T_{\rm eff}$-axis.

\begin{deluxetable*}{rrrrrrrrrhhrrrrrrrrr}
		\tablecaption{Basic stellar informatin, spectral feature measurements, and derived quantities of YSO$_{\rm RF}$ candidates \label{tab:properties}}
		\tablewidth{700pt}
		\tabletypesize{\scriptsize}
		\tablehead{
			\colhead{RA} & \colhead{DE} & \colhead{Plx}  & \colhead{RUWE} & \colhead{Jmag} & \colhead{Nspec} & 
			\colhead{EW6563} & \colhead{EW6708} &  \colhead{EW7065} & \nocolhead{EW8190} &  \nocolhead{EW8542} &  \colhead{...} &
			\colhead{$T_{\rm eff}$}    & \colhead{$\log {\rm t}$}  & \colhead{log$g$}  & \colhead{Mass} & \colhead{$\log \dot{M}_{\rm acc}$}  & \colhead{CTTS?}\\ 
			\colhead{(deg)} & \colhead{(deg)} & \colhead{(mas)} & \colhead{} & \colhead{(mag)} & \colhead{} & 
			\colhead{(\AA)} & \colhead{(\AA)} & \colhead{(\AA)} & \nocolhead{(\AA)}  & \nocolhead{(\AA)}   & \colhead{...} & 
			\colhead{(K)}  &\colhead{} &\colhead{} &\colhead{($M_{\odot}$)} & \colhead{}  & \colhead{} 
		} 
		\startdata		
 28.345498   & 39.122285   &  1.4281  & 1.105  &  14.577  &   3  &    0.169  &  -0.118  &  -0.062  &  -1.167  &  -1.992 & ...  &    3779  & 7.528  & 4.47  & 0.69    &         &   \\
 45.587656   & 17.176216   &  4.0628  & 3.428  &  10.670  &   2  &   88.508  &  -0.344  &   0.114  &  -1.362  &  -0.262 & ...  &    3592  & 5.944  & 3.49  & 0.41    &  -8.119 & yes \\
 76.664548   & -2.643738   &  2.6413  & 1.049  &  13.779  &   2  &    9.100  &   0.152  &  -0.391  &  -0.440  &  -0.338 & ...  &    3222  & 7.178  & 4.25  & 0.41    &         &       \\
 77.023759   & -2.790594   &  2.5430  & 1.030  &  13.665  &   2  &   56.106  &  -0.629  &   0.042  &  -0.410  &  -0.250 & ...  &    3232  & 7.068  & 4.18  & 0.42    &  -9.968 & yes \\
 77.264325   & -2.220214   &  3.0242  & 1.034  &  13.659  &   2  &    4.854  &   0.131  &  -0.248  &  -0.405  &  -0.430 & ...  &    3184  & 7.236  & 4.28  & 0.38    &         &       \\
100.143839   &  9.826120   &          &        &  14.682  &   2  &   32.200  &   0.054  &   0.027  &  -1.557  &  -1.960 & ...  &    3827  &        &       &         &         & yes  \\         
106.220100   & 11.851919   &  3.6647  & 1.608  &  11.209  &   1  &    0.807  &  -0.242  &  -0.113  &  -1.393  &  -1.631 & ...  &    4062  & 6.620  & 4.07  & 0.73    &         &  \\
		\enddata
	\tablecomments{Table~\ref{tab:properties} is published in its entirety in the machine-readable format. A portion is shown here for guidance regarding its form and content.}
	\end{deluxetable*}

\subsection{Accretion signatures of disk-bearing candidates}\label{sec:macc}
Having determined the basic properties, we now focus on constraining the accretion signatures for disk-bearing stars such as CTTS candidates. In the current understanding of magnetospheric accretion, gas from the inner disk edge accretes onto the stellar surface, guided by magnetic field lines in accretion columns or funnel flows. The infalling gas is heated to temperatures of about 8000 K or higher, producing excess continuum emission and broad emission lines \citep[see, e.g., the review by][]{Hartmann2016}. Consequently, several emission lines carry information on the mass accretion rate ($\dot{M}_{\rm acc}$) from the disk onto the star.

\begin{figure*}
	\gridline{\fig{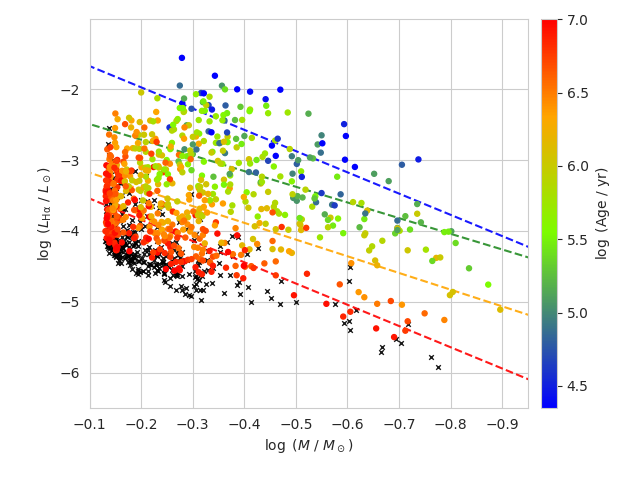}{0.5\textwidth}{}
		\fig{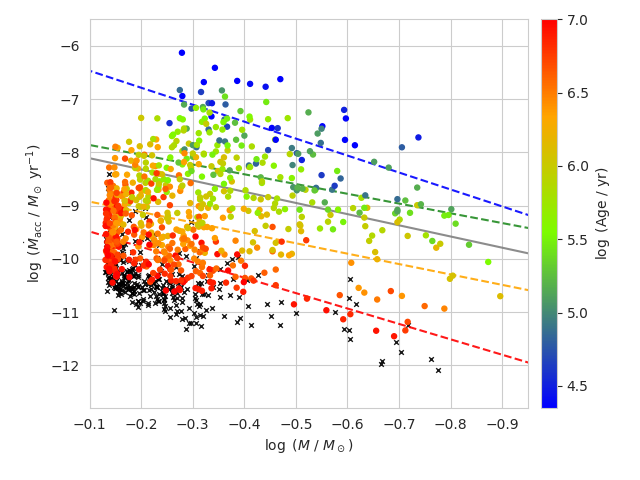}{0.5\textwidth}{}}
	\caption{Left: the $\rm H\alpha$ luminosity of CTTS candidates likely younger than 50 Myr. Circles (color-coded by age) represent those younger than 10 Myr, while black crosses represent those aged 10-50 Myr. The blue dashed line ($\log L_{\rm H\alpha} = -1.37 + 3.00 \times \log M_{\star}$), green dashed line ($\log L_{\rm H\alpha} = -2.27 + 2.21 \times \log M_{\star}$), orange dashed line ($\log L_{\rm H\alpha} = -2.94 + 2.36 \times \log M_{\star}$), and red dashed line ($\log L_{\rm H\alpha} = -3.24 + 3.00 \times \log M_{\star}$) are linear fits to targets with ages of $<$0.1 Myr, 0.1-1 Myr, 1-5 Myr, and 5-10 Myr, respectively. 
	Right: same as the left panel, but for the accretion rate $\dot{M}_{\rm acc}$ derived from $\rm H\alpha$ luminosity. The gray solid line denotes the relation $\dot{M}_{\rm acc} \varpropto M^{2.1}_{\star}$ provided by~\citet{Hartmann2016}. The blue dashed line ($\log \dot{M}_{\rm acc} = -6.15 + 3.18 \times \log M_{\star}$), green dashed line ($\log \dot{M}_{\rm acc} = -7.68 + 1.83 \times \log M_{\star}$), orange dashed line ($\log \dot{M}_{\rm acc} = -8.73 + 1.95 \times \log M_{\star}$), and red dashed line ($\log \dot{M}_{\rm acc} = -9.20 + 2.89 \times \log M_{\star}$) are linear fits to targets with ages of $<$0.1 Myr, 0.1-1 Myr, 1-5 Myr, and 5-10 Myr, respectively.  \label{fig:mass_lha_macc}}
\end{figure*}

In this study, $\dot{M}_{\rm acc}$ was estimated from accretion luminosity ($L_{\rm acc}$), using the following equation \citep[e.g.][]{Gullbring1998}:
\begin{equation} 
	{\dot{M}}_{\rm acc} \simeq \left(1-\frac{R_{\star}}{R_{\rm in}}\right)^{-1}\frac{L_{\rm {acc}} R_{\star}}{GM_{\star}}  \simeq 1.25\frac{L_{\rm acc} R_{\star}}{GM_{\star}},
\end{equation}
where $M_{\star}$ and $R_{\star}$ are the stellar mass and radius, respectively, and $R_{\rm in}$ is the inner disk radius. Here $R_{\rm in} = 5R_{\star}$ \citep{Gullbring1998}, considering that the inner disk radius should be less than the co-rotation radius (typically 5-6 $R_{\star}$) to allow magnetospheric accretion down to the star. The accretion luminosity $L_{\rm acc}$ was estimated from the H$\alpha$ emission luminosity ($L_{\rm H\alpha}$) using the relationship provided by \citet{Herczeg2008}, i.e., $\log(L_{\rm acc}/L_{\odot}) = 2.0 + 1.20 \times \log(L_{\rm H\alpha}/L_{\odot})$. The $L_{\rm H\alpha}$ was calculated using the following formula:
\begin{equation} \label{equ:lline_lamost}
	L_{\rm H\alpha} = {\rm EW^{'}_{H\alpha}} \times F_{c}(\rm H\alpha) \times 4\pi R_{\star}^2, 
\end{equation}
where $\rm EW^{'}_{H\alpha}$ is the H$\alpha$ excess equivalent width, obtained by subtracting the basal value of M dwarfs with similar spectral types ($\rm EW^{'}_{H\alpha} = EW6563 - EW6563_{basal}$). This step removes the contribution of photospheric absorption and any systematic offset due to continuum-band selection. $F_{c}(\rm H\alpha)$ is the stellar surface continuum flux of H$\alpha$ line for the star with the same temperature and surface gravity, evaluated using PHOENIX version 15.5 model spectra (BT-Settl/CIFIST2011\_2015) with solar metallicity \citep{Allard2012,Baraffe2015} and given $\log g$ estimated from {\sc parsec} models. The derived $L_{\rm H\alpha}$ for CTTS candidates younger than 50 Myr are displayed in the left panel of Figure~\ref{fig:mass_lha_macc}, and the corresponding $\dot{M}_{\rm acc}$ values are shown in the right panel. 

It is critical to recognize that LAMOST spectra are calibrated for relative flux rather than absolute flux \citep{Luo2015}, which limits our ability to derive $L_{\rm H\alpha}$ through the potentially more precise method of using the absolute local continuum flux around H$\alpha$ line in conjunction with the stellar distance. The inherent uncertainties associated with $R_{\star}$ and $F_{c}(\rm H\alpha)$ may introduce unpredictable uncertainties or even systematic biases into our determination of $L_{\rm H\alpha}$. Within the current scope of this work, the present analysis is restricted to evaluating internal uncertainties inherent to our $L_{\rm H\alpha}$ derivation methodology rather than making external calibration comparisons. The uncertainty in $L_{\rm H\alpha}$ primarily propagates from the uncertainties in $T_{\rm eff}$ (also assuming $\sigma(T_{\rm eff}) = 200 $K, which mainly affects the selection of $F_{c}(\rm H\alpha)$ values), EW6563 (its error was estimated from the observed flux error, typically $\lesssim0.2$ \AA), and $R_{\star}$ (with a typical uncertainty value of 0.08 $R_{\odot}$, as mentioned above). A Monte Carlo simulation indicates that the uncertainty in $\log (L_{\rm H\alpha}/L_{\sun})$ has a typical value of about 0.23, dominated by the uncertainty in $T_{\rm eff}$ mainly due to its significant effect on $F_{c}(\rm H\alpha)$. The uncertainties in $\dot{M}_{\rm acc}$ are also considered, as they arise from the uncertainties in $L_{\rm H\alpha}$ (thus ${L}_{\rm acc}$), stellar mass and radius. A similar Monte Carlo analysis indicates that the uncertainties in $\dot{M}_{\rm acc}$ are dominated by those in $L_{\rm H\alpha}$, with a typical value of $\log \dot{M}_{\rm acc}$ uncertainty around 0.29.

\section{Discussion}
\subsection{The complexity of Lithium depletion in low-mass YSOs}
The efficient convective mixing effect in PMS stars ensures that surface material is thoroughly mixed with the interior. When this material reaches depths where temperatures are high enough to initiate Li burning, the surface Li abundance decreases rapidly. The time required to reach the Li-burning temperature in the cores of contracting PMS stars is highly sensitive to mass, leading to a complex, mass-dependent pattern of Li depletion in groups of low-mass, coeval stars \citep[e.g.,][]{Jeffries2023}. Therefore, historically, Li has served as an intrinsic age indicator for young stars, particularly in young open clusters \citep[e.g.,][]{Jeffries2009,Randich2021}.

\begin{figure*}
	\gridline{
		\fig{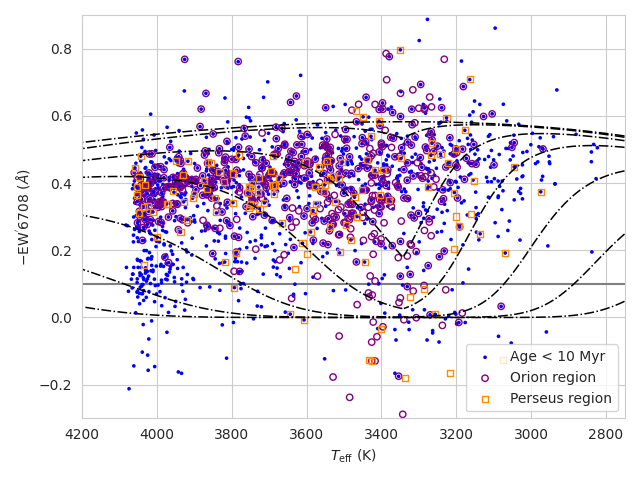}{0.5\textwidth}{}
		\fig{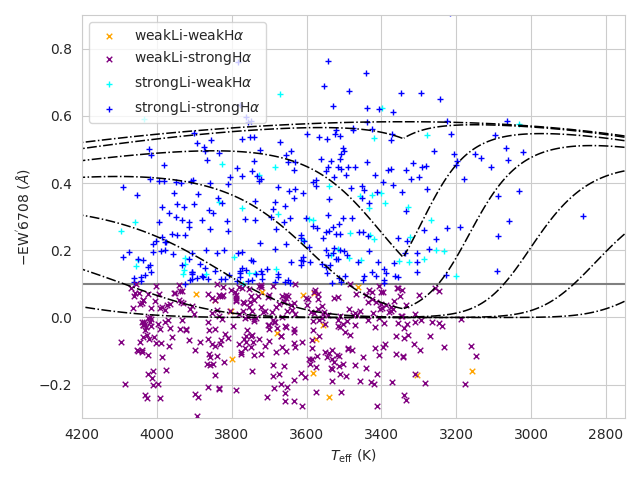}{0.5\textwidth}{}}
	\caption{Left: the excess equivalent width ($\rm EW^{'}6708$) of {Li \sc i} $\lambda6708$ \AA~plotted against $T_{\rm eff}$ for $\rm YSO_{RF}$ candidates potentially younger than 10 Myr, superimposed with $\rm YSO_{RF}$ candidates in the Orion and Perseus star-forming regions. $\rm EW^{'}6708=EW6708-EW6708_{basal}$ represents the difference between the observed value (EW6708) of a given target and the reference value (EW6708$_{basal}$) of dwarfs/giants with similar spectral types. Since we define the equivalent width of an absorption line as negative, the y-axis is plotted with $\rm -EW^{'}6708$ value, allowing for comparison with the model values from the literature. The black dotted-dashed lines represent empirical models of the equivalent width of {Li \sc i} $\lambda6708$ \AA~line (from top to down: 5, 10, 15, 20, 30, 50 and 100 Myr) provided by \citet{Jeffries2023}. The solid gray line at $\rm EW^{'}6708 = -0.1$~\AA, corresponds to the black solid line in panel (a) of Figure~\ref{fig:cuts}, serving as a Li-filter for identify $\rm {YSO}_{Li}$ candidates. Right: same as left panel, but for $\rm YSO_{RF}$ candidates with $\rm EW^{'}6731 > 2$~\AA. These candidates are divided into four groups based on the intensity of the {Li \sc i} $\lambda6708$ \AA~absorption and H$\alpha$ emission, as described in section~\ref{sec:valid2}.
		\label{fig:cmd_li_valid}}
\end{figure*}

A significant feature of this pattern is that some low-mass stars deplete almost all their lithium during the PMS stage. For instance, at an age of about 20 Myr, the surface Li abundance of a PMS star with $T_{\rm eff} \sim 3300$~K drops to levels comparable to those of dwarfs with similar temperatures. By 50 Myr, most PMS stars with $T_{\rm eff}$ in the range of $3100-3800$ K have essentially depleted their lithium \citep[][see also Figure~\ref{fig:cmd_li_valid}]{Jeffries2023}. Furthermore, severe Li depletion has been observed in some very young, low-mass stars with ages of only a few Myr \citep[e.g.,][]{Sacco2008, McBride2021}. This may be attributed to episodic accretion during their early evolution stages \citep[e.g.,][]{Baraffe2010}. The complexity of the Li depletion pattern suggests that any simple Li-filter identification method may miss potential YSOs.

Figure~\ref{fig:cmd_li_valid} shows the distribution of equivalent width of the {Li \sc i} $\lambda6708$ \AA~line for $\rm {YSO}_{RF}$ candidates likely younger than 10 Myr. These stars, with $T_{\rm eff}$ around 3300 K,
exhibit weaker Li absorptions than expected for their ages, with some showing undetectable Li absorptions. Additionally, over half of weak-Li-strong-H$\alpha$ $\rm {YSO}_{RF}$ candidates exhibit signs of star-disk interaction environments (see Section~\ref{sec:valid2}). This indicates that these stars have likely depleted their surface Li at very early ages, possibly younger than 10 Myr, considering that the typical lifetime of circumstellar disks is $\sim10$ Myr \citep[e.g.][]{Fedele2010,Ribas2014,Venuti2019} (although disks around stars in sparse associations may evolve more slowly than those in more crowded environments like clusters \citep[e.g.,][]{Fang2013}). Therefore, the {Li \sc i} $\lambda6708$ \AA~line is no longer a reliable tool for identifying these young stars.
 
\subsection{Mass accretion rates in CTTSs}
In this section, we examine the mass accretion rates in our sample of stars with potential disks. It has been established that the mass accretion rate correlates with stellar mass across the entire range from brown dwarfs to T Tauri stars, following the relation $\dot{M}_{\rm acc} \propto M^{\alpha}_{\star}$, where $\alpha$ ranges from 1.5 to 3.1 \citep[e.g.][and references therein]{Hartmann2016}. The right panel of Figure~\ref{fig:mass_lha_macc} illustrates the derived $\dot{M}_{\rm acc}$ as a function of stellar mass, color-coded by age, for CTTS candidates likely younger than 50 Myr. The dashed lines indicate that the estimated $\alpha$ falls within the range of 1.8 to 3.2 for different sub-age groups, consistent with previous findings.

Figure~\ref{fig:mass_lha_macc} also demonstrates that younger stars generally exhibit higher $\dot{M}_{\rm acc}$, as expect. Both theoretical models of disk evolution and observational data indicate a decline in accretion rate with age \citep[e.g.][]{Hartmann1998,Antoniucci2014}, described by $\dot{M}_{\rm acc} \propto t^{\alpha}$, where $\alpha$ ranges from $-1.6$ to $-1.1$ \citep[see a review by][]{Hartmann2016}. To investigate this relationship, we performed linear fits to the mass accretion rate as a function of age for individual sub-samples with similar masses (to decouple the mass dependence) based on the sample shown in Figure \ref{fig:mass_lha_macc}. The results, as listed in Table~\ref{tab:accretionfits}, yield similar $\alpha$ values ranging from $-1.5$ to $-1.2$. 

The derived accretion rates in Figure~\ref{fig:mass_lha_macc} exhibit significant scatter (e.g., typically  $\ga 1$ dex in $\log \dot{M}_{\rm acc}$) for stars with similar masses, even at similar ages, e.g., a phenomenon observed in previous studies \citep[e.g.,][]{Alcala2014}. Potential sources of this scatter include errors in estimating $\dot{M}_{\rm acc}$ and temporal variability in line emission. The typical error in $\log \dot{M}_{\rm acc}$ is approximately 0.29 dex, as discussed in section~\ref{sec:macc}, which is notably smaller than the observed scatter. Therefore, accretion variability likely contributes significantly to the scatter. Indeed, accretion variability is common among CTTSs \citep[e.g.,][]{Stauffer2014, Venuti2014}, but it seems that intrinsic variability alone cannot fully explain the observed scatter in accretion rates for stars of similar mass \citep[e.g.,][]{Nguyen2009, Fang2013}.  

\begin{figure}	
	\gridline{\fig{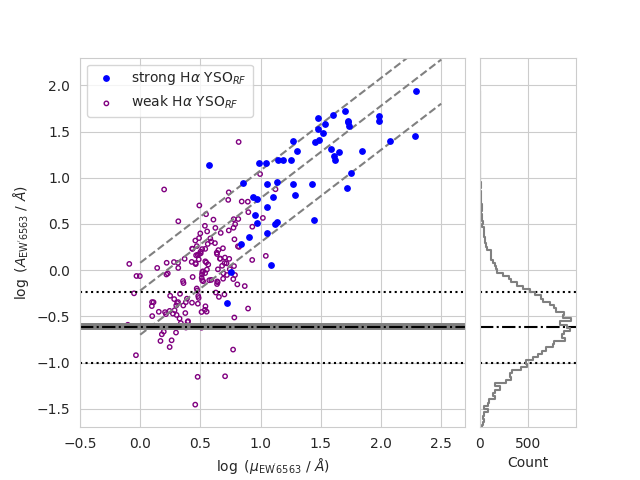}{0.5\textwidth}{}}
	\caption{Peak-to-peak amplitudes ($A_{\rm{EW^{'}6563}}$) versus mean values ($\mu_{\rm{EW^{'}6563}}$) of the equivalent widths of the H$\alpha$ line for $\rm YSO_{RF}$ candidates with at least three valid epochs of spectra. Open purple circles denote those with weak H$\alpha$ emissions, while blue dots represent those with strong H$\alpha$ emissions (CTTS candidates). The dash-dotted line represents the 50th percentile of $A_{\rm{EW6563}}$ for all 24,512 targets in $\rm S_{0}$ sample with at least three valid epochs of spectra. The two dotted black lines denote the 16th and 84th percentiles, as shown in the histogram in the right marginal panel. The solid, bold gray line indicates $A_{\rm{EW6563}}=0.25$~\AA, determined by a Monte Carlo simulation with a typical measurement error of $\sigma_{\rm EW6563} = 0.15$ \AA. The simulated value aligns well with the 50th percentile, suggesting that the observed variability in EW6563 for half of the targets is primarily due to flux uncertainties. The three dashed gray lines (from bottom to top) represent constant ratios of amplitudes to mean values, corresponding to relative changes (the ratio of half amplitude to the mean value) of 10\%, 30\%, and 60\% , respectively. 
		\label{fig:ampha}}
\end{figure}

\begin{figure*}
	\gridline{\fig{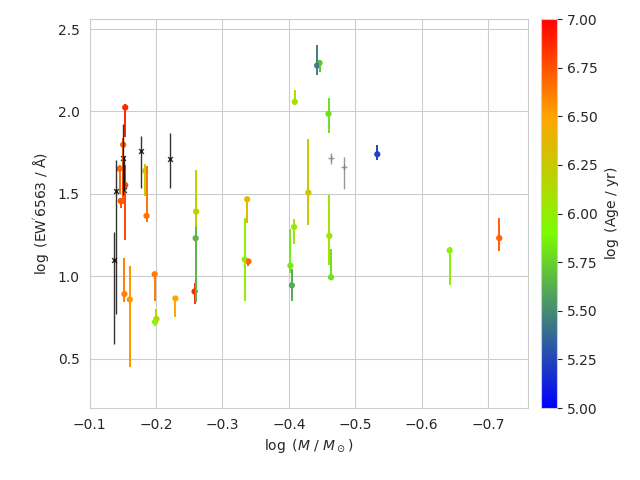}{0.5\textwidth}{}
		\fig{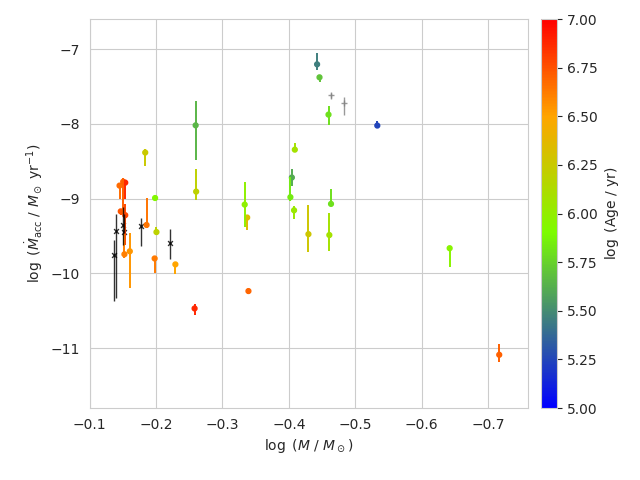}{0.5\textwidth}{}}
	\caption{Variability in the equivalent width of H$\alpha$ emission (left) and mass accretion rate (right) for CTTS candidates likely younger than 50 Myr with at least three epochs of spectra. Symbols (color-coded circles, black crosses, and gray pluses) have the same meanings as in Figure~\ref{fig:mass_lha_macc}, representing the median values of multi-epoch observations for each target, with bars indicating the range of peak-to-peak variation.   \label{fig:ampmacc}}
\end{figure*}

Figure~\ref{fig:ampha} shows the peak-to-peak amplitude of H$\alpha$ equivalent widths for $\rm YSO_{RF}$ candidates with at least three epochs of observations. The data reveal significant temporal variability in H$\alpha$ emission, particularly for CTTS candidates, with typical amplitudes of about 10 \AA~and relative changes (a ratio of the amplitude to the mean value) of about 30\%. The left panel of Figure~\ref{fig:ampmacc} displays the variability of H$\alpha$ equivalent widths as a function of stellar mass for CTTS candidates younger than 50 Myr, while the right panel shows the corresponding variability in derived accretion rates. The typical peak-to-peak amplitude in $\log \dot{M}_{\rm acc}$ is $\sim0.3$ dex. More massive stars appear to exhibit slightly greater variability compared to less massive stars of similar age, but there is no clear dependence of variability on age among stars of similar mass. Combined, measurement uncertainties and temporal variability contribute approximately 0.6 dex to the scatter in $\log \dot{M}_{\rm acc}$, and it does not fully explain the observed scatters. 

CTTSs frequently exhibit spectral veiling effects, a phenomenon first introduced by \citet{Joy1949}. This spectral distortion primarily arises from accretion-driven excess emission in both continuum and line features. The continuum component is attributed to the shock-heated photosphere and pre-shock as well as attenuated post-shock regions \citep{Calvet1998}, while line veiling may originate from post-shock gas \citep{Dodin2012} and/or infalling pre-shock material within accretion columns \citep{SiciliaAguilar2015}. Although we do not delve into the nature of veiling in this work, we assess its implications for determining mass accretion rates. 

A fundamental consideration involves veiling's impact on H$\alpha$ equivalent width measurements. Continuum veiling reduces the observed $\rm H\alpha$ equivalent width, necessitating a correction factor of ($1+r_{\rm c}$), where $r_{\rm c}$ quantifies the continuum veiling level. Adopting a characteristic $r_{\rm c}\sim 0.5$ around $\rm H\alpha$ line for our CTTS candidates, as frequently observed in low-mass CTTSs \citep[e.g.,][]{Hartigan1989, Nelissen2023}, our previously derived mass accretion rates should be systematically adjusted upward by about 0.18 dex in $\log \dot{M}_{\rm acc}$ to account for this effect. Moreover, veiling varies among stars with different mass accretion rates \citep[e.g.,][]{Nelissen2023} and may exhibit temporal variability due to the rotation modulation of the accretion shock, potentially introducing additional scatter in the measured mass accretion rates.
 
Furthermore, veiling-induced suppression of atomic and molecular absorption features (e.g., TiO bands) introduces systematic biases in effective temperature determinations. To quantitatively assess this effect, synthetic veiling experiments were conducted by superimposing excess blackbody continua (with $T_{\rm eff}\sim6000$ K) on several reference spectra of M dwarfs and giants with spectral subtype from M0 to M4. These preliminary simulations reveal that a veiling value of $r_{\rm c} \sim 0.5$ near H$\alpha$ could elevate the measured $T_{\rm eff}$ by 100--300 K, with the magnitude contingent upon spectral subtype and surface gravity. However, it should be noted that magnetic cool spots may conversely strengthen molecular absorption features through localized temperature contrasts \citep[e.g.,][]{Fang2016,GullySantiago2017,PerezPaolino2024}, potentially counteracting some of the veiling-induced rise in effective temperature. Given the competing effects of accretion-generated hot spots and magnetic cool spots on spectral features, we conservatively retain original $T_{\rm eff}$ measurements for subsequent $\dot{M}_{\rm acc}$ estimations. This approach presumes minimal net temperature bias under typical CTTS conditions, though we explicitly recognize this assumption as an unresolved systematic uncertainty that will require quantification through future multi-component spectral modeling efforts.

Conducting complex time-domain spectroscopic studies of CTTSs, such as investigating the temporal variations in H$\alpha$ emission (whether they arise from changes in accretion activity, variations in chromospheric activity levels including flares, or simply due to rotational modulation), and examining the temporal characteristics of veiling, as well as refining corresponding veiling correction to the observational data, is necessary and effective for understanding the scatter in accretion rates. However, the detailed properties of the temporal variation of accretion rates are beyond the scope of this work and will be addressed in future studies, where we will investigate the variability patterns of H$\alpha$ emission and even the veilings in low-mass YSOs, particularly using LAMOST time-domain medium-resolution spectra.

\begin{deluxetable}{lcccccc}
	\tablecaption{Results of linear fits to mass accretion rate as a function of age with $\log \dot{M}_{\rm acc} = {\rm intercept + slope} \times \log t$  \label{tab:accretionfits}}
	\tablewidth{0pt}
	\tablehead{
		\colhead{Mass range}  & \colhead{slope} & \colhead{intercept} & \colhead{$r$-value}  &  \colhead{$p$-value}  & \colhead{counts} & \colhead{$\log t$ range}  
	}
	\startdata
	0.10 - 0.15 & -1.19 & -3.42 & -1.00 & 3.4e-02 & 3  & 5.32 - 6.12 \\
	0.15 - 0.20 & -1.44 & -1.32 & -0.95 & 1.5e-13 & 26 & 4.48 - 7.24 \\
	0.20 - 0.25 & -1.19 & -2.62 & -0.91 & 1.0e-12 & 32 & 4.31 - 7.20 \\
	0.25 - 0.30 & -1.34 & -1.57 & -0.91 & 6.4e-13 & 32 & 4.36 - 7.05 \\
	0.30 - 0.35 & -1.17 & -2.22 & -0.85 & 6.4e-13 & 43 & 4.30 - 7.07 \\
	0.35 - 0.40 & -1.30 & -1.16 & -0.84 & 3.4e-13 & 46 & 4.17 - 7.02 \\
	0.40 - 0.45 & -1.45 & -0.07 & -0.90 & 7.8e-21 & 56 & 4.30 - 7.21 \\
	0.45 - 0.50 & -1.42 & -0.34 & -0.90 & 5.0e-34 & 93 & 4.02 - 7.28 \\
	0.50 - 0.55 & -1.49 & 0.26  & -0.89 & 6.6e-29 & 80 & 4.17 - 7.22 \\
	0.55 - 0.60 & -1.34 & -0.75 & -0.79 & 3.7e-20 & 88 & 4.56 - 7.27 \\
	0.60 - 0.65 & -1.43 & 0.01  & -0.76 & 1.5e-15 & 75 & 5.80 - 7.28 \\
	0.65 - 0.70 & -1.29 & -0.72 & -0.69 & 6.0e-09 & 54 & 5.98 - 7.30 \\
	0.70 - 0.75 & -1.32 & -0.55 & -0.62 & 1.9e-18 & 163 & 6.22 - 7.29 \\
	\enddata
\end{deluxetable}

\section{Summary}
This study utilized LAMOST low-resolution spectra to identify low-mass YSOs and characterize their accretion signatures. We investigated the spectral features of different M-type stars (YSOs, dwarfs, and giants) and focused on red-band features that effectively distinguish between these classes. In addition to the commonly used {Li \sc i} $\lambda6708$ \AA~absorption and H$\alpha$ emission lines, other features such as {He \sc i}, {Na \sc i}, {Ca~\sc ii} H\&K and IRT lines, and molecular bands of CN, CaH, and VO showed potential for identifying YSOs from M dwarfs and giants.
Using key red-band features -- H$\alpha$, {Li \sc i} $\lambda6708$ \AA, {Na \sc i} doublets, Ca~{\sc ii} $\lambda8542$ \AA, He~{\sc i} $\lambda7065$ \AA, and molecular bands of CaH3 and VO1 -- we identified over 8,500 M-type YSO candidates from the LAMOST DR8 low-resolution spectra archive using the random forest method, with over 2,300 of them likely being CTTS candidates. These M-type YSO candidates are predominantly located in well-known star-forming regions, such as the Orion and Perseus star-forming regions.

Basic properties, including mass, radius, and age, were estimated using {\sc parsec} evolutionary models with data from $Gaia$ astrometry, 2MASS photometry, and 3D dust maps. For CTTS candidates, accretion-powered H$\alpha$ emissions were converted to mass accretion rates. The derived accretion rates range from $\sim10^{-11}$ to $\sim10^{-7}$ ${M}_{\odot}$~yr$^{-1}$ and depend on both stellar mass and age. Significant scatter in mass accretion rates exists even among stars with similar age and mass. The combination of measurement uncertainties and temporal variability in H$\alpha$ emission does not fully explained these observed scatter, suggesting that other factors, such as veiling due to accretion, are also at play.

The complexity of lithium depletion in low-mass YSOs was preliminarily investigated using our YSO candidates. Our results show that many YSOs exhibit weaker Li absorptions than expected for their ages, with some showing undetectable Li. This suggests that these stars deplete their surface Li at very early ages, possibly younger than 10 Myr. Consequently, the {Li \sc i} $\lambda6708$ \AA~absorption is no longer a reliable tool for identifying such young stars. 

It is important to note that the output of the random forest algorithm depends on the training sample and input variables. Our experience confirms that the final identified YSO sample varies based on the training set and spectral features used. However, the main conclusions of this study remain robust and unaffected by these issues.

\begin{acknowledgments}
We would like to express our sincere gratitude to the anonymous reviewer for the insightful feedback and valuable suggestions, which have significantly enhanced the rigor and clarity of our manuscript. This work was supported by the National Natural Science Foundation of China (NSFC) under grant Nos. 12090040/4, 12373036, 12073047. The Guo Shou Jing Telescope (the Large sky Area Multi-Object fiber Spectroscopic Telescope, LAMOST) is a National Major Scientific Project built by the Chinese Academy of Sciences. Funding for the project has been provided by the National Development and Reform Commission. LAMOST is operated and managed by National Astronomical Observatories, Chinese Academy of Sciences. This work has made use of data from the European Space Agency (ESA) mission {\it Gaia} (\url{https://www.cosmos.esa.int/gaia}), processed by the $Gaia$ Data Processing and Analysis Consortium (DPAC, \url{https://www.cosmos.esa.int/web/gaia/dpac/consortium}). Funding for the DPAC has been provided by national institutions, in particular the institutions participating in the {\it Gaia} Multilateral Agreement. This research has made use of NASA's Astrophysics Data System (ADS) Abstract Service, and of the VizieR catalogue access tool and the cross-match service provided by CDS, Strasbourg, France.
\end{acknowledgments}

%

\vspace{5mm}


\software{TOPCAT \citep{Taylor2005}, NumPy \citep{Walt2011}, Matplotlib \citep{Hunter2007}, Astropy \citep{AstropyCollaboration2013,AstropyCollaboration2018}, Scikit-learn \citep{Pedregosa2012}}



\appendix
\section{Training samples of M-type YSOs, dwarfs, and giants} \label{sec:knownyso}
To effectively guide our YSO classification process and to train the Random Forest classifier (see Appendix~\ref{sec:rfc}), we included both known M-type YSOs and non-YSO stars in our sample. This involved searching not only the SIMBAD database but also the relevant literature to compile a comprehensive list of known YSOs and non-YSOs. These samples were classified into three categories: YSOs, dwarf stars, and giant stars. Their spectra were retrieved from the LAMOST DR8 low-resolution archive.

\subsection{M-type YSOs} 
We gathered low-mass YSO candidates from the SIMBAD database using the `maintype' keyword (e.g., maintype = YO, TT, pr*, Or*, and FU*). Additionally, we utilized several catalogs from the literature to collect known YSOs located in well-known star-forming regions, such as the Taurus-Auriga Complex \citep{Rebull2010,Luhman2010,Dzib2015}, Orion complex \citep{Fang2009,Megeath2012,SzegediElek2013,kim2016,Furlan2016,Ansdell2017}, Perseus, and other regions \citep{Young2015,Luhman2016,Getman2017}. Initially, we retrieved over 1,100 spectra with SNR$r$ greater than 10 for approximately 840 late-K and M-type YSO candidates from the LAMOST DR8 low-resolution archive. When possible, we excluded potentially non-YSO candidates classified based on IR features by \citet{Marton2019} (those accepted as YSO candidates with a probability of less than 50\%). At this stage, 928 spectra (693 targets) remained in our list. Among these YSO candidates, about 30 targets exhibited H$\alpha$-absorption-like features (e.g., located on the dwarf and even giant branches in the EW6563 vs. TiO5n diagram, see panel (d) of Figure~\ref{fig:cuts}). These were further removed, leaving a final sample of 881 spectra of 655 M-type training YSOs.
\subsection{M dwarfs} 
To compile a training sample of M-type dwarfs, we began with the M dwarf catalog identified by \citet{Zhong2019} with LAMOST spectra, which are located in the dwarf branch of the well-known TiO5 vs. [CaH] (= CaH2+CaH3) diagram \citep[e.g.][]{Mann2012,Lepine2013}. We retrieved over 510,000 spectra with ${\rm SNR}_r>10$ from the LAMOST DR8 archive for approximately 460,000 confirmed M dwarfs in the catalog. To remove contamination from YSOs and evolved stars, when available, we excluded targets whose `maintype' in SIMBAD was YSO-like (e.g., YO, TT, pr*, Or*, and FU*) or giant-like (e.g., RGB, AGB, post-AGB, and Mira), or those with a probability of being YSOs greater than 40\% as provided by \citet{Marton2019}. At this stage, approximately 95\% of the spectra remained. We further refined the sample by selecting higher SNR$_r$ for early-subtype M dwarfs, e.g., ${\rm SNR}_r > 60$ for M0-M1, ${\rm SNR}_r > 50$ for M2, ${\rm SNR}_r > 35$ for M3, ${\rm SNR}_r > 25$ for M4, and ${\rm SNR}_r > 15$ for others, leaving a sample of approximately 70,000 spectra of M dwarf candidates. Since it was unnecessary to use all of them to train the classifier, we randomly selected about 5,100 spectra, ensuring a similar number of spectra for each subtype. Finally, we removed a few spectra with EW6563 $>10$~\AA~or EW6563 less than typical values for giants, resulting in a final training sample of 5,084 spectra of 4,909 M dwarf candidates.
\subsection{M giants} 
Following a similar procedure to that used for selecting M dwarfs, we primarily compiled the M giant training sample based on the M giant catalog of \citet{Zhong2019}. We retrieved approximately 43,000 spectra with ${\rm SNR}_r > 10$ from the LAMOST DR8 archive for about 35,000 confirmed M dwarfs in the catalog. To remove contamination from YSOs, when available, we excluded targets whose `maintype' in SIMBAD was YSO-like or those with a probability of being YSOs greater than 40\% as provided by \citet{Marton2019}. At this stage, 41,800 spectra remained. We further refined the sample by selecting higher SNR$_r$ for early subtype M giants, leaving a sample of approximately 30,000 spectra of M giants. We then randomly selected about 5,000 spectra from these over 30,000 spectra, ensuring a similar number of spectra for each subtype. Finally, after removing spectra with strong H$\alpha$ emissions, our final training sample included 4,868 spectra of 4,560 M giant candidates.

\subsection{M-type member candidates in $\alpha$ Per, Pleiades, and Hyades} \label{sec:ocs}
For comparison, we also retrieved LAMOST DR8 spectra with ${\rm SNR}_r > 10$ of the M-type members in three nearby, young, benchmark open clusters: $\alpha$ Per~\citep[$\sim$90 Myr; ][]{Stauffer1999}, Pleiades~\citep[$\sim$120 Myr;][]{Stauffer1998, Dahm2015}, and Hyades~\citep[$\sim$700 Myr;][]{Perryman1998, Douglas2019}. Specifically, we obtained 105 valid spectra of 95 M-type $\alpha$ Per members based on the membership provided by \citet{Lodieu2019}; 1,118 valid spectra of 362 Pleiades member candidates, and 428 spectra of 218 Hyades members, following the member-selection procedures of \citet{Fang2018}.

\section{The potentially informative spectral features} \label{sec:features}
\begin{figure*}
\plotone{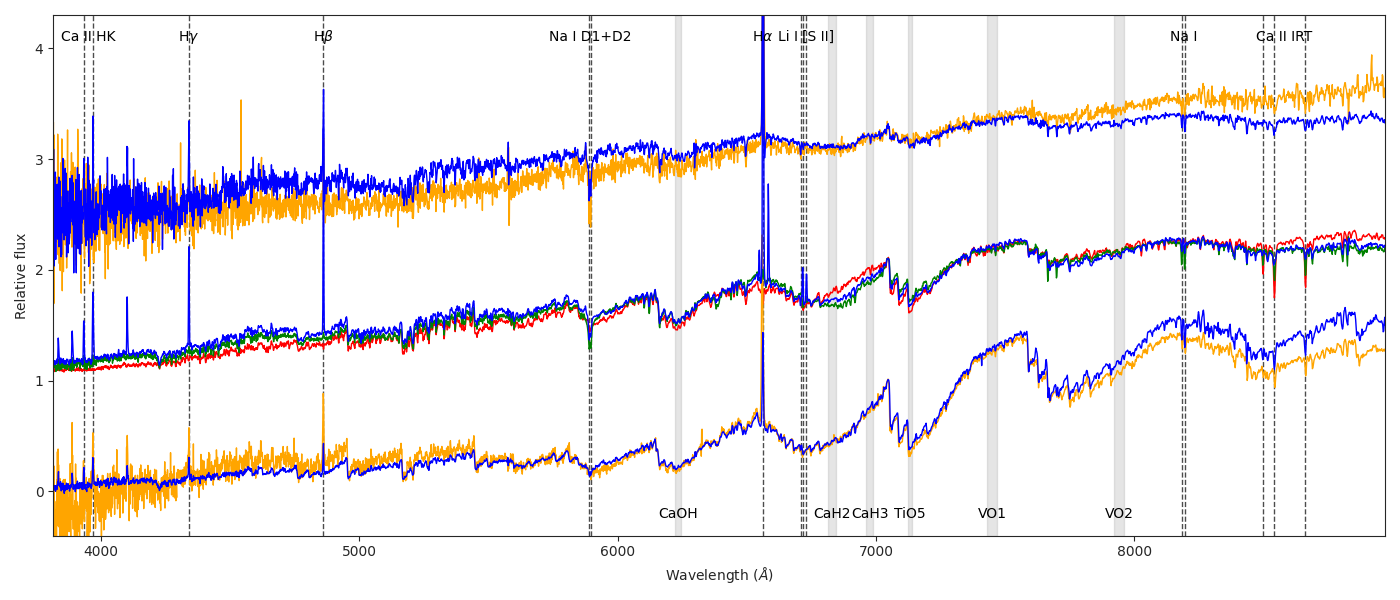}
	\caption{Representative spectra of M-type stars, with each spectrum normalized to the flux at the TiO band head near 7050~\AA~and vertically offset for clarity. The top displays two spectra of an M0 YSO candidate, collected at different epochs and exhibiting different signal-to-noise ratios (blue: ${\rm SNR}_r\sim43$; orange: ${\rm SNR}_r\sim11$). The bottom showcases two spectra from an M4 YSO, also obtained at different epochs and having varying qualities (blue: ${\rm SNR}_r\sim70$; orange: ${\rm SNR}_r\sim11$). Centrally located are exemplary spectra of M2-type stars (all have ${\rm SNR}_r>65$), featuring a YSO (blue), a dwarf (green), and a giant (red). For reference, key spectral features under analysis are indicated.
		\label{fig:example}}
\end{figure*}

Spectroscopically identifying YSO candidates among unknown sources is more challenging than anticipated. Each YSO exhibits unique spectral features due to the complex structure of the star-disk system and processes such as inflow and outflow, compounded by temporal variability. Therefore, relying on just one or two features is insufficient for accurate identification. Instead, multiple informative features are required to robustly identify YSO candidates. To achieve this, we measured nearly 30 potentially informative spectral features, as detailed in Tables~\ref{tab:index} and~\ref{tab:width}. Figure \ref{fig:example} showcases representative spectra for M-type stars with subtypes M0, M2, and M4, with prominent spectral features that were analyzed clearly marked. 

\subsection{Metrics of spectral features}
The energy distribution of M-type stars significantly deviates from that of a blackbody, characterized by numerous strong absorptions of molecular bands, such as CaOH ($\sim$6230~\AA), CaH (e.g., $\sim$6800 and 6960~\AA), and TiO (e.g., $\sim$6160, 7055, and 7126\AA), as shown by Figure~\ref{fig:example}. We quantified these molecular bands using spectral indices, defined as the ratio of the average flux within a characteristic band (numerator) to that in a nearby pseudo-continuum (denominator). The definitions of these indices are listed in Table~\ref{tab:index}, with the exception of S4142, which is defined as the ``magnitude" of the total flux in the band relative to that in the nearby pseudo-continuum, describing the strength of the CN molecular band near 4215~\AA~\citep{Smith1996}.

In addition to well-defined molecular absorption features, the spectra of M-type young stars also exhibit numerous absorption and/or emission lines, including Hydrogen Balmer lines, Ca~{\sc ii} H\&K, Ca~{\sc ii} IRT lines, and several neutral metal lines such as Ca~{\sc i} $\lambda 4227$, Mg~{\sc i} b triplet, and Na~{\sc i} D lines. Some YSOs also show emissions in forbidden lines, such as [O~{\sc i}], [S~{\sc ii}], and [N~{\sc ii}]. We measured the equivalent widths of various characteristic lines, with definitions listed in Table~\ref{tab:width}. For example, we measured the equivalent width of the line H$\alpha$ (EW6563) using the following formula:
\begin{equation} 
	\label{equ:ewide}
	\rm{EW} = \int \frac{F_{\lambda}-F_{c}}{F_{c}}d\lambda,
\end{equation}
where the line is centered at 6563~\AA~over a 14~\AA~bandpass, and the continuum flux $F_{c}$ is the average flux between 6520--6540 and 6595--6615~\AA. A positive equivalent width value indicates emission within this metric. Some of our measurements are displayed in Figure~\ref{fig:cuts}, and Figure~\ref{fig:caiik} to~\ref{fig:molecular}.

It is important to note that for M-type spectra dominated by molecular absorptions, it is challenging to identify the ``real" nearby continuum for many spectral features. Therefore, all measurements were obtained directly from the non-normalized observed spectrum, and the continuum bands used were fixed for all subtypes. As a result, the intrinsic behavior of a given feature across different spectral types may be mixed with changes in continuum flux, such as variations in molecular absorption strength. For instance, the rapid rise of EW8542 in M-type giants after $\rm TiO5n < 0.4$, as shown by Figure~\ref{fig:cuts}, is primarily due to the rapid changes in the nearby `continuum' flux as molecular absorption bands appear. Similar behaviors are observed in other spectral features, such as EW6708 and EW8190.

\subsection{Measurement uncertainties}
The uncertainties in the equivalent width and spectral index measurements primarily arise from the flux errors inherent in the spectra. The uncertainty at each wavelength pixel in the LAMOST spectra, encompassing Poisson noise and any noise introduced during data reduction, is documented and has been directly propagated to the corresponding measurements. The estimated measurement errors due to flux uncertainties are listed in Table~\ref{tab:properties} (available in the online version).

As expected, the lower signal-to-noise ratio in the blue wavelength region leads to higher uncertainties in the measurements of features compared to those in the red band. For example, for spectra with a typical ${\rm SNR}_r \sim 20$, the representative errors in the equivalent widths are 0.15~\AA~for H$\alpha$ (EW6563), 0.55~\AA~for H$\beta$ (EW4861), and 1.15~\AA~for H$\gamma$ (EW4340). Notably, the measurement of {Ca \sc ii} K line (EW3934) is particularly affected by significant noise, resulting in a typical error of approximately 2.8~\AA.  In contrast, the equivalent width of the {Ca \sc ii} infrared triplet line (EW8542) exhibits much lower errors, with a typical value of 0.11~\AA. Given the considerable uncertainties associated with certain blue-band features, our analysis prioritizes spectral features in the red wavelength region, where higher SNR generally yields more reliable measurements.

\begin{deluxetable*}{lllll}
	\tablecaption{Definition of spectral indices\label{tab:index}}
	\tablewidth{0pt}
	\tablehead{
		\colhead{Index} & \colhead{Numerator (\AA)} & \colhead{Denominator (\AA)} & \colhead{Feature} & \colhead{Reference}  
	}
	\startdata
	S4142      & 4120--4216     & 4216--4290             &  CN band  &  \citet{Smith1996} \\
	CH4300     & 4300--4309     & 4316--4321             &  G band   &  this work\\
	CaOH       & 6230--6240     & 6345--6354             & CaOH band & \citet{Reid1995} \\
	CaH2n      & 6814--6846     & 7042--7048             &  CaH band &  this work \\
	CaH3n      & 6960--6990     & 7042--7048             &  CaH band &  this work \\
	TiO5n      & 7126--7135     & 7042--7048             &  TiO band &  \citet{Fang2016} \\
	VO1        & 7430--7470     & 7550--7570             &  VO band  & \citet{Lepine2003} \\
	VO2        & 7920--7960     & 8130--8150             &  VO band  & \citet{Lepine2003} \\
	\enddata
\end{deluxetable*}

\begin{deluxetable*}{lllll}
	\tablecaption{Definition of equivalent widths\label{tab:width}}
	\tablewidth{0pt}
	\tablehead{
		\colhead{EW} & \colhead{Line bandpass (\AA)} & \colhead{Pseudo-continua (\AA)} & Feature  
	}
	\startdata
	EW3934    &  3930--3937     &  3905--3920, 3945--3960 & Ca~{\sc ii} K                      \\    
	EW4340    &  4336--4345     &  4317--4327, 4370--4380 & H$\gamma$                          \\
	EW4861    &  4855--4867     &  4842--4852, 4873--4883 & H$\beta$                           \\    
	EW5018    &  5015--5022     &  5010--5015, 5025--5030 & Fe~{\sc ii} $\lambda5018$          \\ 
	EW5876    &  5872--5879     &  5830--5850, 5910--5930 & He~{\sc i} $\lambda5876$ (D$_{3}$) \\
	EW5890    &  5885--5900     &  5830--5850, 5910--5930 & Na~{\sc i} D (D$_{1}$+D$_{2}$)     \\
	EW6300    &  6295--6305     &  6260--6275, 6320--6335 & [O~{\sc i}] $\lambda6300$      \\
	EW6363    &  6359--6367     &  6320--6335, 6385--6400 & [O~{\sc i}] $\lambda6363$      \\
	EW6548    &  6544--6552     &  6520--6540, 6595--6615 & [N~{\sc ii}] $\lambda6548$     \\
	EW6563    &  6556--6570     &  6520--6540, 6595--6615 & H$\alpha$                      \\
	EW6583    &  6579--6587     &  6520--6540, 6595--6615 & [N~{\sc ii}] $\lambda6583$     \\
	EW6678    &  6675--6681     &  6650--6670, 6685--6705 & He~{\sc i} $\lambda6678$       \\
	EW6708    &  6705--6710     &  6694--6704, 6710-6712  & Li~{\sc i} $\lambda6708$       \\
	EW6716    &  6712--6721     &  6685--6705, 6745--6765 & [S~{\sc ii}] $\lambda6716$     \\
	EW6731    &  6724--6738     &  6685--6705, 6745--6765 & [S~{\sc ii}] $\lambda6731$     \\
	EW7065    &  7062--7068     &  7055--7061, 7072--7080 & He~{\sc i} $\lambda7065$       \\
	EW8190    &  8176--8202     &  8165--8175, 8208--8218 & Na~{\sc i} doublets              \\
	EW8498    &  8493--8503     &  8477--8487, 8520--8530 & Ca~{\sc ii} IRT ($\lambda8498$)  \\
	EW8542    &  8536.5--8547.5 &  8520--8530, 8555--8565 & Ca~{\sc ii} IRT ($\lambda8542$)  \\
	EW8662    &  8656.5--8667.5 &  8640--8650, 8676--8686 & Ca~{\sc ii} IRT ($\lambda8662$)  \\
	\enddata                                                    
\end{deluxetable*}                                       

\subsection{An overview of measured spectral features}\label{sec:overview}
YSOs are nascent stars, often possess circumstellar disks, accretion-powered winds, and outflowing jets, as well as strong magnetic fields. These phenomena leave distinct imprints on their optical spectra.

Lithium (Li) is a fragile element that is destroyed at relatively low temperatures ($\sim$2.5$\times$10$^{6}$ K) via proton capture reactions. As YSOs evolve towards the ZAMS, Li in the atmosphere begins to deplete when core temperatures become sufficiently high, making lithium a powerful age indicator for young, cool stars \citep[e.g.,][]{Dahm2008,Fang2009}. A hallmark feature of YSOs is the strong absorption of the {Li \sc i} resonance line at about 6708~\AA. As shown by panel (a) in Figure~\ref{fig:cuts}, M-type YSOs typically exhibit stronger {Li \sc i} $\lambda6708$ \AA~absorptions compared to most M-type giants and dwarfs, including stars near the ZAMS, such as those in the Pleiades.

\begin{figure*}
	\gridline{\fig{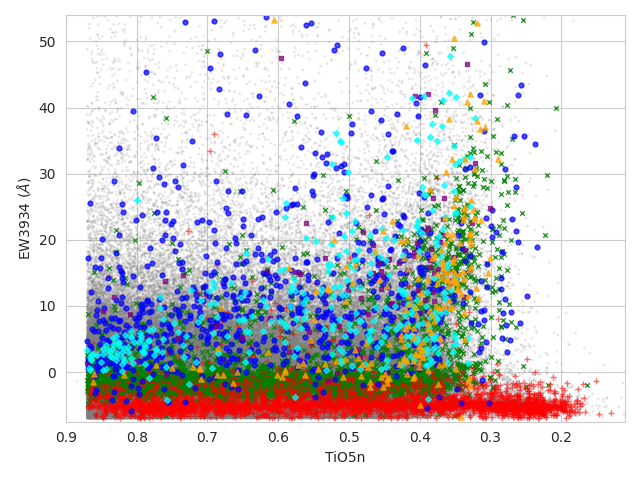}{0.5\textwidth}{}
		\fig{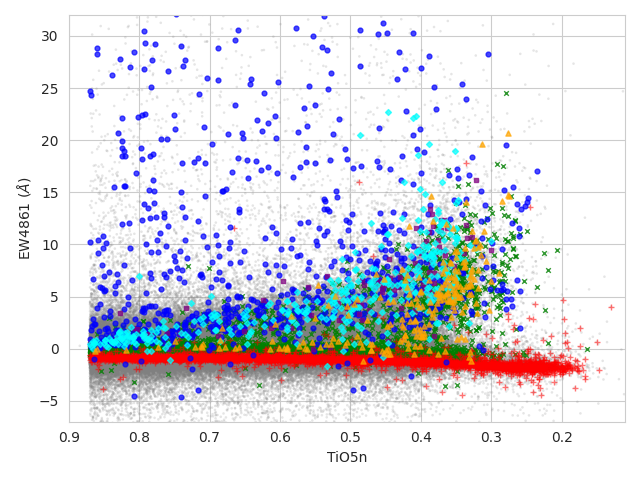}{0.5\textwidth}{}}
	\caption{The equivalent widths of {Ca \sc ii} K (EW3934, left) and $\rm H\beta$ (EW4861, right) lines plotted against TiO5n. The symbols are the same as those in Figrue~\ref{fig:cuts}. 
	\label{fig:caiik}}
\end{figure*}

Young M dwarfs, such as those in the Pleiades, Hyades, and Praesepe, show strong chromospheric emissions in Hydrogen Balmer and {Ca \sc ii} H\&K lines \citep[e.g.,][]{Fang2018}. YSOs often display similar emissions due to their magnetically active chromospheres. Additionally, accretion-powered emissions contribute to these lines. As illustrated in panel (d) of Figure~\ref{fig:cuts} and Figure~\ref{fig:caiik}, YSOs exhibit strong emissions in H$\alpha$, $\rm H\beta$ and {Ca \sc ii} K lines, often surpassing the emissions seen in Pleiades members. The {Ca \sc ii} IRT lines are also used as indicators of chromospheric activity and can show emissions due to magnetospheric infall of gas from the disk \citep{Azevedo2006, Connelley2010}. Panel (b) of Figure~\ref{fig:cuts} shows the equivalent width of {Ca \sc ii} $\lambda8542$ \AA~line highlighting the significant excess emissions in YSOs compared to dwarfs and Pleiades members with similar spectral types.

\begin{figure*}
	\gridline{\fig{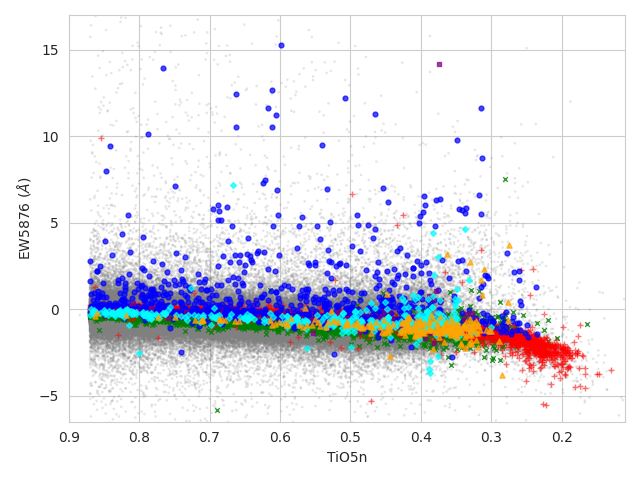}{0.5\textwidth}{}
		\fig{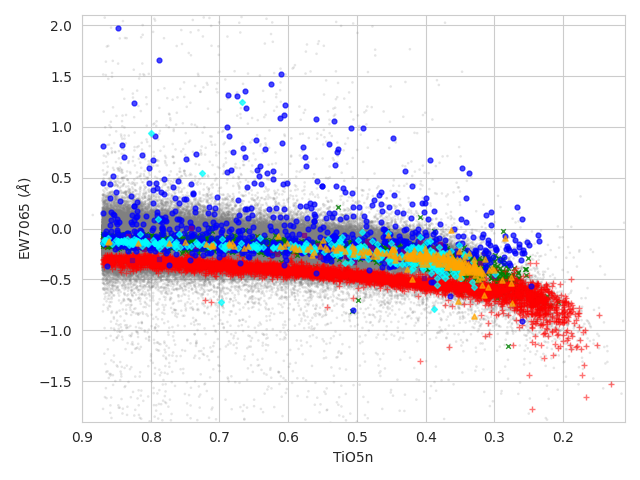}{0.5\textwidth}{}}
	\caption{Same as Figure~\ref{fig:caiik}, but for {He \sc i} $\lambda5876$ and $\lambda7065$ \AA~lines.
	\label{fig:hei}}
\end{figure*}

Low-mass YSOs, particularly CTTSs, often show prominent {He \sc i} emission lines, which originate primarily from accretion columns and/or hot winds \citep[e.g.,][]{Beristain2001}. We measured the equivalent widths of {He \sc i} lines at 5876, 6678, and 7065~\AA~and detected evident emissions in these lines among many YSOs. Figure~\ref{fig:hei} displays the measured values of {He \sc i} $\lambda5876$ and {He \sc i} $\lambda7065$ \AA~showing that many YSOs exhibit emissions relative to dwarfs and giants. It is worth noting that {He \sc i} emissions can also result from flaring activity, with {He \sc i} $\lambda5876$ \AA~originating in the upper chromosphere and serving as an indicator of flare-like events \citep{Montes1996}. Cataclysmic variables often exhibit clear emissions in {He \sc i} lines \citep[e.g.,][]{Han2018, Hou2020}, which can reduce the reliability of using this feature to identify YSOs.

\begin{figure*}
	\gridline{\fig{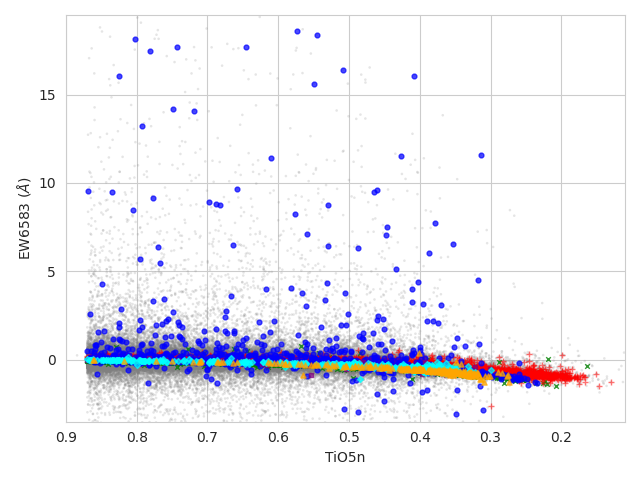}{0.5\textwidth}{}
		\fig{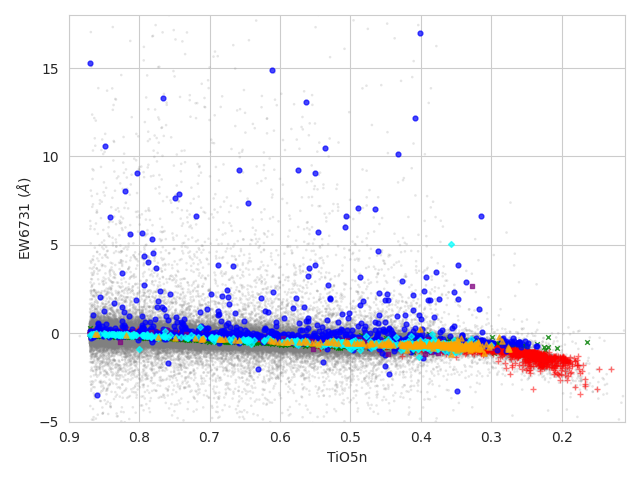}{0.5\textwidth}{}}
	\caption{Same as Figure~\ref{fig:caiik}, but for [N~{\sc ii}] $\lambda6583$ \AA~and [S~{\sc ii}] $\lambda6731$ \AA~lines.
	\label{fig:niisii}}
\end{figure*}

Emissions in forbidden lines such as [{O \sc i}], [{N \sc ii}], and [{S \sc ii}] have been reported among disk-bearing YSOs \citep[e.g.,][]{Hartigan1995, Fang2009}. These emissions are likely from accretion-powered winds and/or outflow jets. Figure~\ref{fig:niisii} shows the measured equivalent widths of [{N \sc ii}] $\lambda6583$ \AA, and [{S \sc ii}] $\lambda6731$ \AA~lines, indicating strong emissions in some YSOs, which suggests active accretion processes.

\begin{figure*}
	\gridline{\fig{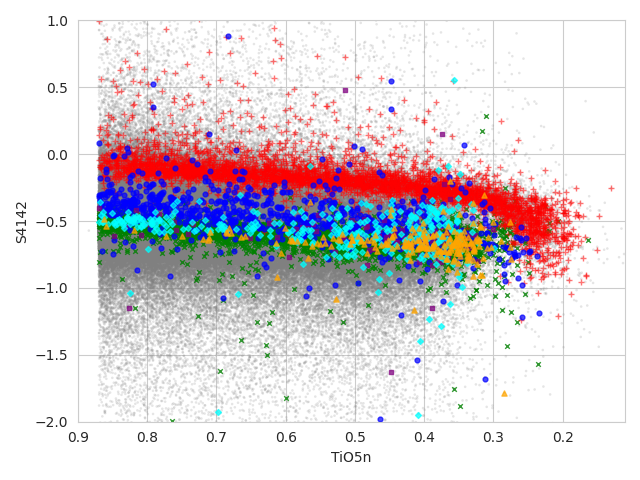}{0.5\textwidth}{(a)}
		\fig{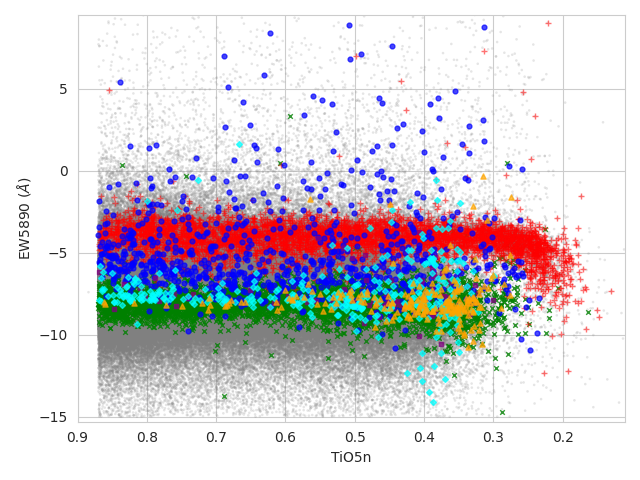}{0.5\textwidth}{(b)}}
	\gridline{\fig{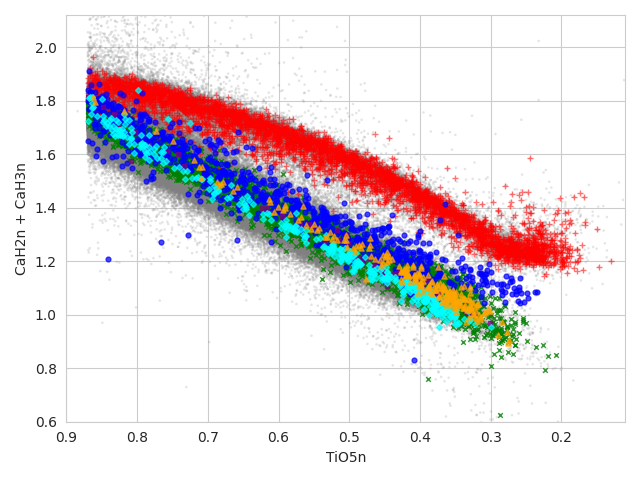}{0.5\textwidth}{(c)}
		\fig{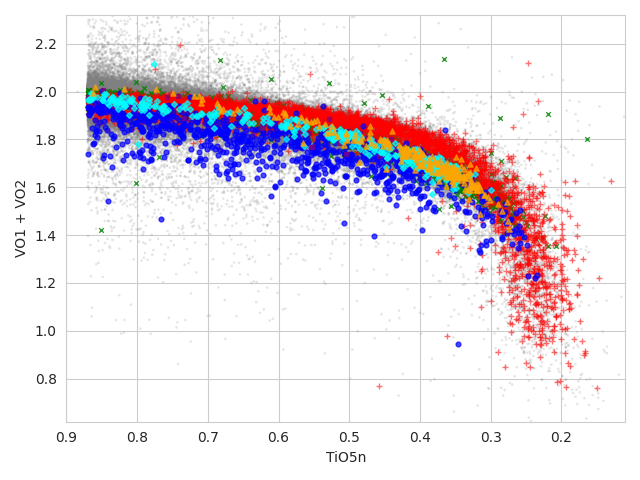}{0.5\textwidth}{(d)}}
	\caption{Same as Figure~\ref{fig:caiik}, but for the equivalent width of {Na \sc i} D lines (EW5890) and the indices of molecular bands of CN (S4142), CaH (CaH2n+CaH3n) and VO (VO1+VO2).
		\label{fig:molecular}}
\end{figure*}

For M-type stars, TiO is sensitive to temperature and metallicity but weakly dependent on gravity \citep{Woolf2006, Lepine2007}, while CaH is sensitive to temperature and gravity but less so to metallicity \citep{Mann2012}. The combination of TiO and CaH serves as a good indicator of gravity, as illustrated in Figure~\ref{fig:molecular}. The different patterns among dwarfs and giants in the [CaH] (CaH2n + CaH3n) vs. TiO5n diagram highlight this distinction. Many other spectral features also show similar gravity/luminosity sensitivities, such as the CN band (e.g., S4142, see panel (a) of Figure~\ref{fig:molecular}), G band (e.g., CH4300), Na {\sc i} lines (e.g., EW8190, see panel (c) of Figure~\ref{fig:cuts}, and EW5890, see panel (b) of Figure~\ref{fig:molecular}), and Ca {\sc ii} IRT lines (e.g., EW8542, see panel (b) of Figure~\ref{fig:cuts}). These features exhibit statistical differences between M giants and dwarfs with similar TiO5n, aiding in the discrimination of M giants from M dwarfs. Pre-main sequence stars have surface gravities similar to cool sub-giants and are offset from dwarf stars \citep{Herczeg2014}, a finding confirmed by our measurements of spectral features such as the CN band (S4142) and Na {\sc i} lines (EW8190), where YSOs tend to locate between dwarfs and giants. However, some features like {Ca \sc ii} IRT lines show different patterns for YSOs due to their strong chromospheric activity and accretion contributions.

\begin{figure*}
	\gridline{\fig{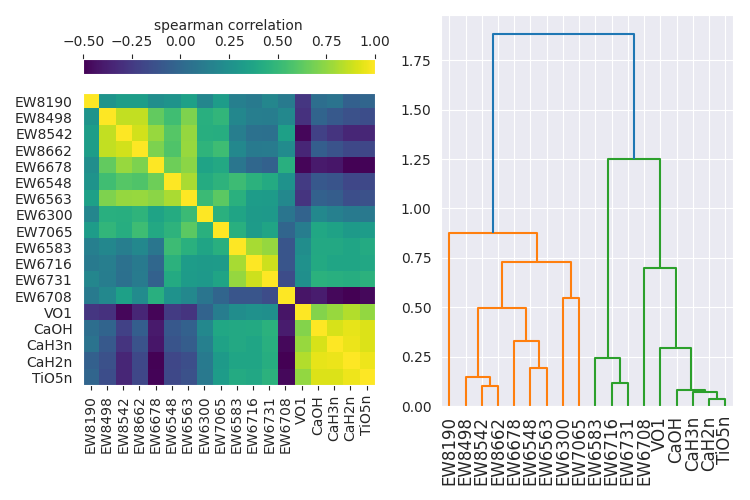}{0.5\textwidth}{(a)}
		\fig{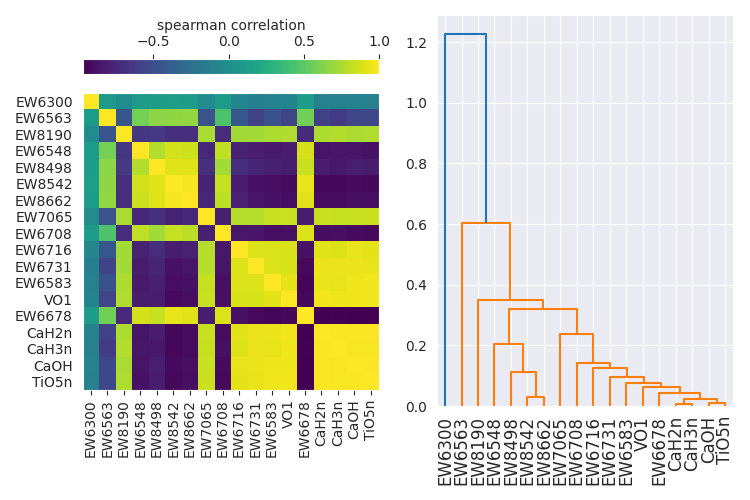}{0.5\textwidth}{(b)}}
	\gridline{\fig{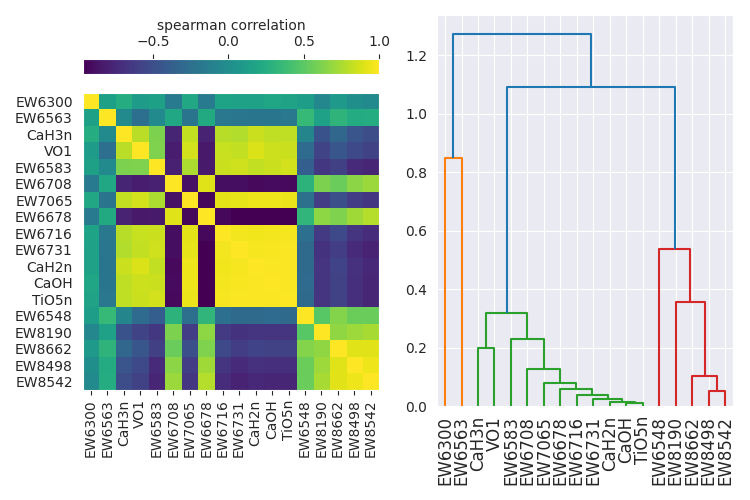}{0.5\textwidth}{(c)}
		\fig{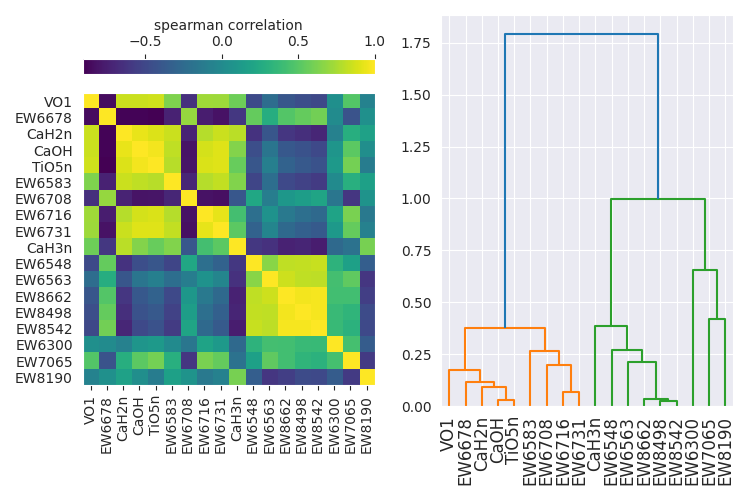}{0.5\textwidth}{(d)}}
	\gridline{\fig{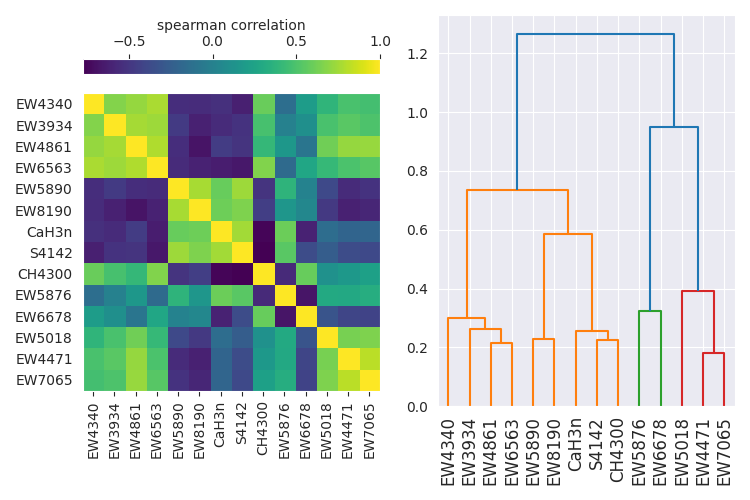}{0.5\textwidth}{(e)}}
	\caption{Spearman rank-order correlations between measurements and the corresponding dendrograms for hierarchical clustering based on these correlations. Panels (a), (b), (c), and (d) show the correlations between features in the red-band for the training sets of YSOs, dwarfs, giants, and the combined set of all three classes, respectively. Panel (e) shows the correlations between blue-band features and selected red-band features for the combined set of all three classes. 
	\label{fig:corr1}}
\end{figure*}

Many spectral features are correlated with each other. We measured the Spearman rank-order correlations between features for stars in the training samples, and performed a hierarchical cluster analysis on these correlations, using the distance matrix derived from the correlation matrix (distance = $1 - |{\rm correlation}|$), and applied Ward's linkage method, as shown in Figures~\ref{fig:corr1}. The results, shown in Figure~\ref{fig:corr1}, illustrate the composition of each cluster by drawing U-shaped links between a cluster and its children. Many features, such as the {Ca \sc ii IRT} lines, exhibit expected correlations with each other. These correlations are useful for guiding and simplifying the selection of formative features in YSO identification (see section~\ref{sec:rfc}).
 
\section{Identify YSO candidates with Random Forest}\label{sec:rfc} 
To identify YSO candidates based on measured spectral features, we employed the Random Forest (RF) algorithm implemented in the {\sc scikit-learn} module \citep{Pedregosa2012}. RF classifiers are ensemble methods that aggregate (majority vote) the classifications of multiple trained decision trees. For a given object, the RF classifier maps sets of spectral features to probabilities for class membership, providing the likelihood that an unclassified object belongs to each class. The method consists of the following three main procedures:

(1) Compilation of labeled training data: The RF algorithm requires a labeled training dataset to infer a function. This step involves compiling a training sample by carefully selecting the most probable YSOs, as well as non-YSOs, including dwarfs and giants. In this work, we classified the training targets into three groups: YSOs, dwarfs, and giants (see Appendix~\ref{sec:knownyso}).

(2) Feature selection and model training: The next step is to identify spectral features that describe the different classes in the most meaningful way and can separate them most efficiently. As discussed in Appendix~\ref{sec:features}, several features are particularly useful for distinguishing these three classes, such as Hydrogen Balmer, {He \sc i}, {Na \sc i}, {Ca \sc ii} HK and IRT lines, CN and CaH bands, and especially the {Li \sc i} $\lambda6708$ \AA~line. Many of these features are correlated with each other, as shown in Figures~\ref{fig:corr1}. From the perspective of the RF classifier, any of these correlated features can serve as predictors without a concrete preference for one over the others. Considering the correlations among these features and the lower noise levels in the red band for M-type stars, we primarily adopted features in the red band to train the RF classifiers. We investigated their efficiencies in separating different classes using the permutation importance technique. This technique measures the decrease in accuracy score when a single feature value is randomly shuffled; the feature with a greater decrease in accuracy generally has higher importance for the trained RF classifier. However, when two features are correlated and one is permuted, the model can still access the feature through its correlated counterpart, leading to lower importance values for both features, even if they are significant. To address this issue, we performed hierarchical clustering on the Spearman rank-order correlations (see Appendix~\ref{sec:features}) and selected one feature from each cluster to ensure that the adopted features used in the permutation importance procedure do not have strong correlations with each other. 

\begin{figure*}[ht!]
	\plotone{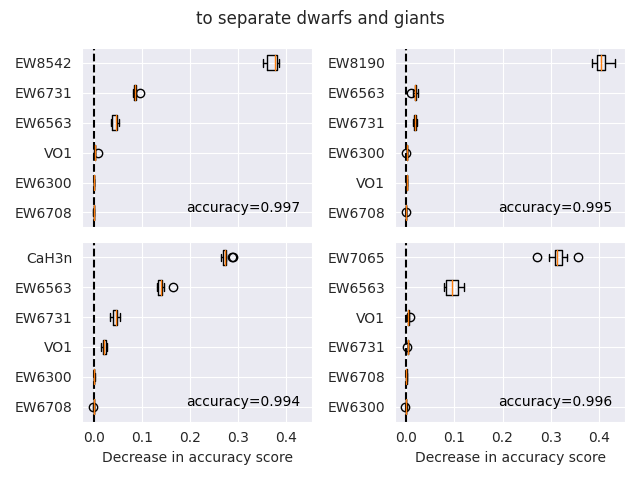}
	\caption{
		Permutation importance of features for a trained RF classifier (for distinguishing dwarfs from giants). This metric reflects the decrease in accuracy when a single feature value is randomly shuffled, indicating the feature's importance to the trained RF model. The permutation importance for each feature was assessed 10 times using cross-validation, involving 10 stratified randomized folds of train-test sets from our training samples. For each panel, 10 RF classifiers were trained with the specified features, yielding 10 values of accuracy decrease for each feature. The box plot in each panel summarizes these values, with the orange line indicating the median. The listed accuracy in each panel represents the mean accuracy score of the corresponding 10 RF classifiers.  
		\label{fig:imp_dg}}
\end{figure*}

\begin{figure*}[ht!]	
	\plotone{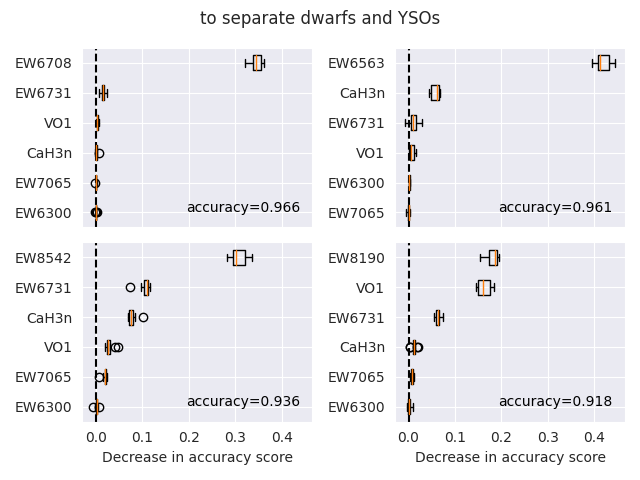}
	\caption{Same as Figure~\ref{fig:imp_dg}, but for distinguishing YSOs from dwarfs. 
		\label{fig:imp_dy}}
\end{figure*}

\begin{figure*}[ht!]
	\plotone{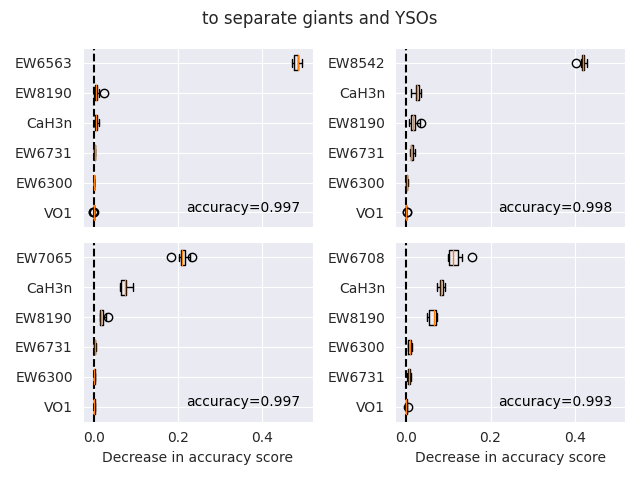}
	\caption{Same as Figure~\ref{fig:imp_dg}, but for distinguishing YSOs from giants.  
		\label{fig:imp_gy}}
\end{figure*}

Figures~\ref{fig:imp_dg} to~\ref{fig:imp_gy} show the permutation importances of several features in the red band for the trained RF classifiers to separate different classes. Key findings include: EW6563, EW8542, EW8190, EW7065, and CaH3n are effective in separating dwarfs and giants; EW6708, EW6563, EW8542, EW8190, and VO1 are highly useful in separating dwarfs and YSOs; EW6563, EW8542, EW7065, EW6708, and CaH3n have high efficiencies in separating giants and YSOs. Finally, we adopted the following features in the red band to train the RF classifier: EW6563, EW6708, EW8190, EW8542, EW7065, CaH3n, and VO1.

(3)Implementation of trained RF classifier: Once the RF classifier is trained, it can be applied to identify the most probable YSOs in the given dataset. The trained RF classifier identified 8,567 YSO candidates in the initial sample $\rm S_0$, which contained over 583,000 M-type targets (see Section~\ref{sec:preselect}).  

\section{Potential systematic uncertainties in age and mass accretion rates}\label{sec:extinction} 
The right panel of Figure~\ref{fig:cmd_parsec} reveals that a significant fraction of YSO candidates exhibit inferred ages exceeding 50 Myr. For instance, within the weak-Li-strong-H$\alpha$ subsample (850 stars with reliable $Gaia$ parameters: $\varpi/\sigma_{\varpi}>5$, $\rm RUWE<1.4$), 389 stars (46\%) are distributed within the 50--200 Myr region of the $M_{\rm J}$--$T_{\rm eff}$ diagram. If these 389 stars are indeed older than 50 Myr, their accretion activities should be minimal or non-existent, given the typical disk lifetime of approximately 10 Myr \citep[e.g.][]{Fedele2010,Ribas2014,Venuti2019}. This expectation conflicts with the observed strong H$\alpha$ emissions and the detection of forbidden line emission ($\rm EW^{'}6731>1$~\AA) in more than half of them. We propose two potential resolutions: (1) systematic underestimation of interstellar extinction ($A_{\rm J}$) may artificially displace these sources toward older isochrones, and/or (2) discrepancies may exist between current evolutionary models and observational constraints. 
	
\begin{figure}
	\gridline{\fig{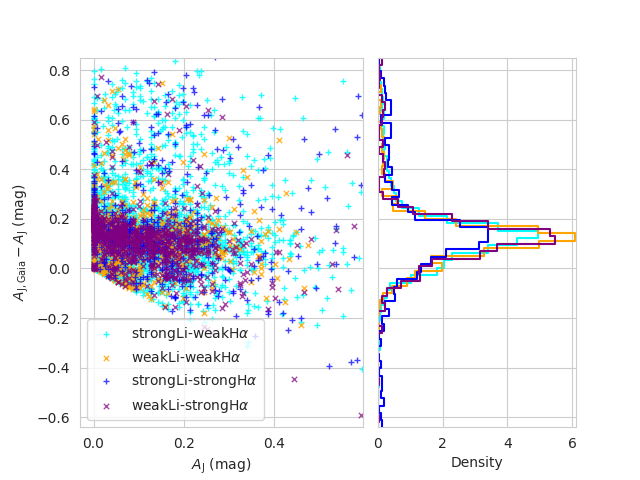}{0.5\textwidth}{}}
	\caption{Comparison of adopted $A_{\rm J}$ values with Gaia DR3 extinction estimates. The $A_{\rm J, Gaia}$ values are derived from $A_{\rm 0}$ values ($Gaia$ DR3; \citealt{Creevey2023,Fouesneau2023}) using a $J$-band extinction coefficient of 0.239 \citep{Danielski2018}. The left panel shows the $\Delta A_{\rm J}$ ($=A_{\rm J, Gaia} - A_{\rm J}$) distribution, with corresponding histograms presented in the right panel.  \label{fig:ajag}}
\end{figure}

Figure~\ref{fig:ajag} compares our adopted $A_{\rm J}$ values with those ($A_{\rm J, Gaia}$) derived from $Gaia$ DR3 extinction \citep{Creevey2023,Fouesneau2023}, revealing a median offset of $\Delta A_{\rm J} (= A_{\rm J, Gaia} - A_{\rm J}) \approx 0.15$ mag (especially when $A_{\rm J} < 0.4$ mag). Although applying a $+0.15$ mag correction to $A_{\rm J}$ would systematically reduce inferred ages by $\Delta \log t \sim -0.1$ dex (equivalent to $\sim20\%$ age reduction), this adjustment alone only reclassifies 16\% of 389 stars as less than 50 Myr. We therefore investigated alternative evolutionary models using the \citet{Baraffe2015} isochrones (hereafter BHAC15), finding systematically younger ages for the subsample with $\Delta \log t$ ranging from $-0.2$ to $-0.6$ dex (median $-0.3$ dex, mass-dependent): 66\% of 389 stars shift to the region younger than 50 Myr. Incorporating both the $A_{\rm J}$ correction and BHAC15 isochrones increases this fraction to 94\%, with nearly 40 cool stars ($T_{\rm eff} < 3600$ K) occupying the region younger than 20 Myr. These results suggest that the apparently old YSO candidates with strong accretion signatures likely represent younger populations when accounting for extinction uncertainties and model-dependent systematics.  

\begin{figure*}
	\gridline{\fig{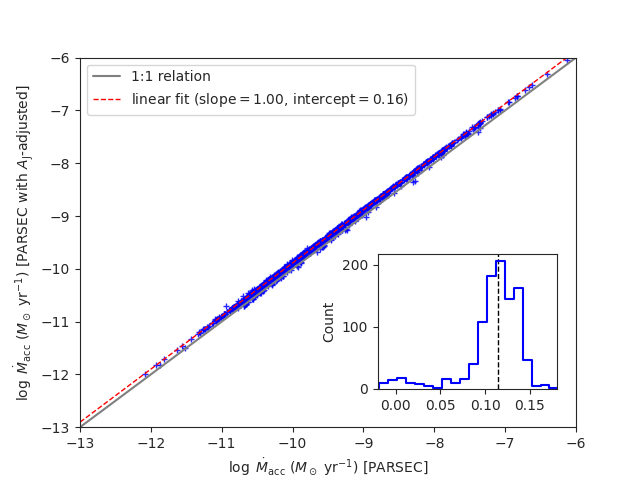}{0.5\textwidth}{}
		\fig{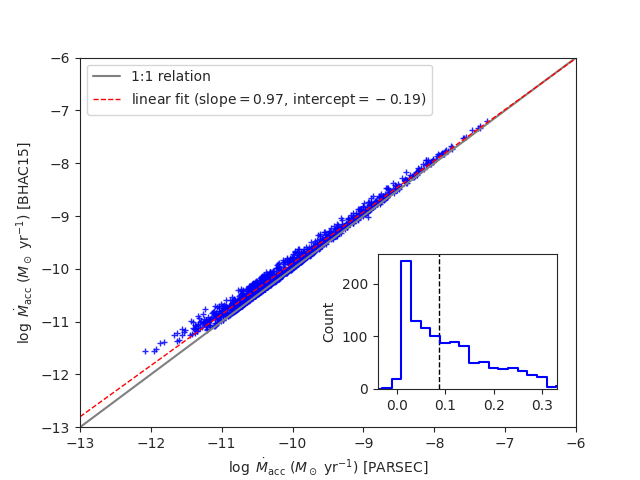}{0.5\textwidth}{}}
	\caption{Comparison of mass accretion rate ($\dot{M}_{\rm acc}$) estimations utilizing different stellar parameters.
		Left: Correlation between $\dot{M}_{\rm acc}$ values estimated based on PARSEC stellar parameters, with and without adjustment for $A_{\rm J}$. The insert panel shows the histogram of the differences, indicating a typical offset in $\log \dot{M}_{\rm acc}$ of $\sim0.11$ dex, as marked by the black dashed line. Right: Discrepancy in $\dot{M}_{\rm acc}$ estimates using BHAC15 and PARSEC stellar parameters (both $A_{\rm J}$-unadjusted). The inset panel highlights the statistical distributions of their differences, showing a median difference in $\log \dot{M}_{\rm acc}$ of $\sim0.09$ dex, as shown by the black dashed line. \label{fig:macc_sys}}
\end{figure*}

\begin{figure}
	\gridline{\fig{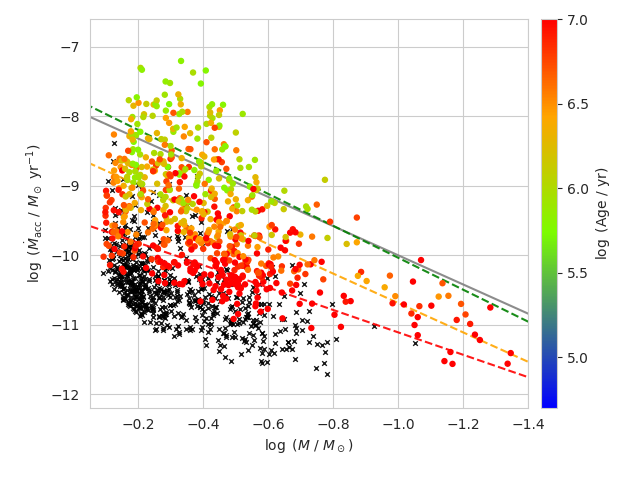}{0.5\textwidth}{}}
	\caption{The accretion rate $\dot{M}_{\rm acc}$ as a function of stellar mass for CTTS candidates younger than 50 Myr, utilizing stellar parameters from the BHAC15 models. Circles (color-coded by age) represent those younger than 10 Myr, while black crosses represent those aged 10--50 Myr. The gray solid line denotes the relation $\dot{M}_{\rm acc} \varpropto M^{2.1}_{\star}$ provided by~\citet{Hartmann2016}. The green dashed line ($\log \dot{M}_{\rm acc} = -7.74 + 2.30 \times \log M_{\star}$), orange dashed line ($\log \dot{M}_{\rm acc} = -8.57 + 2.12 \times \log M_{\star}$), and red dashed line ($\log \dot{M}_{\rm acc} = -9.50 + 1.61 \times \log M_{\star}$) are linear fits to targets with ages of 0.5--1 Myr, 1--5 Myr, and 5--10 Myr, respectively.  \label{fig:mass_macc_bhac15}}
\end{figure}

In light of the above, we examine the systematic uncertainties in mass accretion rate estimates for CTTS candidates arising from these potential discrepancies. The left panel of Figure~\ref{fig:macc_sys} compares the $\dot{M}_{\rm acc}$ values based on stellar parameters from PARSEC models with and without $A_{\rm J}$ adjustments, demonstrating that a $+0.15$ mag adjustment in $A_{\rm J}$ increases $\dot{M}_{\rm acc}$ by $\Delta \log \dot{M}_{\rm acc}\sim0.11$ dex. Comparative analysis in the right panel reveals a 0.09 dex difference between PARSEC and BHAC15 model-derived $\dot{M}_{\rm acc}$ values. Investigation of $\dot{M}_{\rm acc}$--mass relation using BHAC15 models yields consistent slopes (from 1.6 to 2.3) across age subgroups (Figure~\ref{fig:mass_macc_bhac15}), in agreement with previous studies and our previous results. Similarly, temporal decay indices ($\alpha$ in $\dot{M}_{\rm acc} \propto t^{\alpha}$) range from $-1.8$ to $-1.0$, aligning with values in Table~\ref{tab:accretionfits}. In summary, despite potential systemic uncertainties in our estimated mass accretion rates due to unknown discrepancies in extinction and/or stellar evolution models, the primary conclusions in this work remain robust.

\bibliography{yso01ref}{}
\bibliographystyle{aasjournal}



\end{document}